\documentclass[reprint,amsmath,amssymb,aps,superscriptaddress]{revtex4-2}
\pdfoutput=1

\usepackage{algorithmic}
\usepackage{listings}
\usepackage{graphicx}
\usepackage[font={small}]{caption}
\usepackage{subcaption}
\usepackage{textcomp}
\usepackage[dvipsnames]{xcolor}
\def\BibTeX{{\rm B\kern-.05em{\sc i\kern-.025em b}\kern-.08em
    T\kern-.1667em\lower.7ex\hbox{E}\kern-.125emX}}

\newcommand{\zz}{\mathcal{ZZ}}
\newcommand{\R}{R_{\phi}}

\usepackage{color, colortbl}
\definecolor{orange}{rgb}{0.992, 0.553, 0.286}
\definecolor{teal}{rgb}{0.024, 0.694, 0.769}

\usepackage{mathtools}
\usepackage{braket}
\usepackage{float}

\usepackage{qcircuit}
\usepackage{hyperref}
\hypersetup{
  breaklinks=true,
  colorlinks=true,
  allcolors=BlueViolet,
}
\usepackage[capitalise]{cleveref}

\Crefname{figure}{Fig.}{Figs.}
\crefname{equation}{}{}

\DeclarePairedDelimiter\abs{\lvert}{\rvert}

\newcommand{\appsection}[1]{\section{\MakeUppercase{#1}}}

\newcommand{\Yb}{\textsuperscript{171}Yb\textsuperscript{+}}

\def\Superstaq/{\texttt{Superstaq}}
\def\qiskit/{\texttt{qiskit}}
\def\cirq/{\texttt{cirq}}

\newcommand{\haware}{h_A}
\newcommand{\hunaware}{h_U}
\newcommand{\infidelity}{\mathcal{I}_{H}}
\newcommand{\perr}{{\epsilon}}

\begin{document}

\preprint{APS/123-QED}

\title{Noise-Aware Circuit Compilations for a \\ Continuously Parameterized Two-Qubit Gateset}

    \author{Christopher G. Yale}
    \email{cgyale@sandia.gov}
    \thanks{These authors contributed equally to this work.}
    \affiliation{Sandia National Laboratories, Albuquerque, New Mexico 87123}

    \author{Rich Rines}
    \email{superstaq@infleqtion.com}
    \thanks{These authors contributed equally to this work.}
    \affiliation{Infleqtion, Chicago, Illinois 60604, USA}

    \author{Victory Omole}
    \email{superstaq@infleqtion.com}
    \thanks{These authors contributed equally to this work.}
    \affiliation{Infleqtion, Chicago, Illinois 60604, USA}

    \author{Bharath Thotakura}
    \email{superstaq@infleqtion.com}
    \thanks{These authors contributed equally to this work.}
    \affiliation{Infleqtion, Chicago, Illinois 60604, USA}

    \author{Ashlyn D. Burch}
    \affiliation{Sandia National Laboratories, Albuquerque, New Mexico 87123}

    \author{Matthew N. H. Chow}
    \affiliation{Sandia National Laboratories, Albuquerque, New Mexico 87123}
    \affiliation{Department of Physics and Astronomy, University of New Mexico, Albuquerque, New Mexico 87131}
    \affiliation{Center for Quantum Information and Control, University of New Mexico, Albuquerque, New Mexico 87131}

    \author{Megan Ivory}
    \affiliation{Sandia National Laboratories, Albuquerque, New Mexico 87123}  

    \author{Daniel Lobser}
    \affiliation{Sandia National Laboratories, Albuquerque, New Mexico 87123}

    \author{Brian K. McFarland}
    \affiliation{Sandia National Laboratories, Albuquerque, New Mexico 87123}
    
    \author{Melissa C. Revelle}
    \affiliation{Sandia National Laboratories, Albuquerque, New Mexico 87123}

    \author{Susan M. Clark}
    \affiliation{Sandia National Laboratories, Albuquerque, New Mexico 87123}

    \author{Pranav Gokhale}
    \affiliation{Infleqtion, Chicago, Illinois 60604, USA}

\date{\today}

\begin{abstract}
State-of-the-art noisy-intermediate-scale quantum (NISQ) processors are currently implemented across a variety of hardware platforms, each with their own distinct gatesets. As such, circuit compilation should not only be aware of, but also deeply connect to, the native gateset and noise properties of each. Trapped-ion processors are one such platform that provides a gateset that can be continuously parameterized across both one- and two-qubit gates. Here we use the Quantum Scientific Computing Open User Testbed (QSCOUT) to study noise-aware compilations focused on continuously parameterized two-qubit $\zz$ gates (based on the M{\o}lmer-S{\o}rensen interaction) using \Superstaq/, a quantum software platform for hardware-aware circuit compiler optimizations. We discuss the realization of $\zz$ gates with arbitrary angle on the all-to-all connected trapped-ion system. Then we discuss a variety of different compiler optimizations that innately target these $\zz$ gates and their noise properties. These optimizations include moving from a restricted maximally entangling gateset to a continuously parameterized one, swap mirroring to further reduce total entangling angle of the operations, focusing the heaviest $\zz$ angle participation on the best performing gate pairs, and circuit approximation to remove the least impactful $\zz$ gates. We demonstrate these compilation approaches on the hardware with randomized quantum volume circuits, observing the potential to realize a larger quantum volume as a result of these optimizations. Using differing yet complementary analysis techniques, we observe the distinct improvements in system performance provided by these noise-aware compilations and study the role of stochastic and coherent error channels for each compilation choice.

\end{abstract}

\maketitle

\section{Introduction}

The utility of today's quantum computers is fundamentally limited by noise. In order to execute an abstract quantum application on a real quantum computer, it must first be compiled to the discrete set of physical operations available to the hardware. The details of this compilation process can have tremendous impact on the quality of experimental results on a limited and noisy device. Well-optimized quantum compilation has the potential to vastly expand the complexity of quantum applications that can be executed meaningfully on today's hardware.

Currently viable qubit technologies---for example superconducting transmons, trapped ions, neutral atoms, and silicon quantum dots---have their own unique set of control mechanisms, topological constraints, timescales, and physical capabilities and tradeoffs. Computationally, these determine the primitive gateset made available by the device and the corresponding connectivity between qubits on the device. Superconducting qubits, for example, support very fast quantum operations via RF signals, but are typically limited to static 2D connectivity; whereas ions and cold atoms support both movement and longer-range interactions via optical pulses.

The physics behind each qubit modality also imbues it with a unique set of dominant error mechanisms and noise sources. To this end, optimized compilation is most successful when it is carefully tuned to the hardware, taking into account the specific capabilities and constraints of the device with a deep awareness of the possible error mechanisms.

Here, we focus on developing optimizations for a trapped-ion architecture. Trapped-ion approaches are a significant player in the quantum computing landscape providing high-fidelity operations~\cite{egan2021, pino2021}, long-lived coherence~\cite{wang2021}, high degree of connectivity~\cite{monroe2013}, and paths to fault tolerance~\cite{egan2021,erhard2021lq,ryananderson2024,hong2024}. In many cases, at current register sizes, these processors realize complete, or ``all-to-all'', connectivity either through the use of a single chain with shared motional modes~\cite{landsman2019}, or through shuttling operations~\cite{pino2021, kielpinski2002}. Trapped-ion and neutral-atom processors are relatively unique in this regard, and any compiler for such a device should take full advantage of this connectivity since no additional gates are needed to bridge any missing connectivity. Likewise, qubits can be relabeled to take advantage of better-performing gate pairs and/or through the use of virtual SWAPs in protocols such as swap mirroring~\cite{campbell2023, cross2019}. 

The ability to realize a continuously parameterized two-qubit gateset has also become a key feature of trapped-ion processors in the last few years~\cite{nam2020, shaffer2023, moses2023}. In many cases, maximally entangling gates are not required nor desired to realize quantum circuits~\cite{lanyon2011, hashim2022, blekos2024, draper2000}. As smaller-angle entanglers tend to incur less error than maximally entangling gates~\cite{ruzic2024, moses2023, perez2023, hashim2022}, this continuous range of entangling angles helps to reduce the overall error of a circuit. Here, we develop a suite of compilation optimizations in \Superstaq/~\cite{campbell2023,infleqtion2024} harnessing the flexibility that is gained by these two features, all-to-all connectivity and continuously parameterized two-qubit gates. We then demonstrate and examine their role in improving performance on a trapped-ion system.

\section{Quantum Scientific Computing Open User Testbed}
\label{sec:qscout}
The experiments here are performed on a trapped-ion quantum processor, the Quantum Scientific Computing Open User Testbed (QSCOUT) at Sandia National Laboratories~\cite{clark2021}. In this work, the qubit register consists of a linear chain of either 4 or 5 ions with all-to-all connectivity. Physical gates are executed using Raman transitions on a hyperfine ``clock'' transition of \Yb~qubits with a pulsed 355~nm laser. The native gateset consists of a continuously parameterized single-qubit phased rotation gate, $\R(\theta)=e^{-i\phi Z/2}e^{i\theta X/2}e^{i\phi Z/2}$ (with Pauli operators $X$, $Z$, rotation angle $\theta$, and phase $\phi$), and a continuously parameterized two-qubit gate, $\zz(\theta) = e^{i\theta Z\otimes Z/2}$. Single-qubit phase gates can be realized via virtual $Z(\theta)$ gates, which shift the phase of all subsequent waveforms. 

For single-qubit gates, $\R(\theta)$, the rotation angle $\theta$ is realized by the pulse duration, and the phase $\phi$ is the programmed relative phase between both legs of the Raman transition. These gates are constructed by putting both tones of the Raman transition on the qubit's respective individual addressing (IA) beam, a co-propagating and motional-insensitive configuration. Additionally, the single-qubit gates are run in parallel and mitigate first-order crosstalk through unique single-photon detunings~\cite{chow2024}. 

The two-qubit gate, based on a M{\o}lmer-S{\o}rensen (MS) interaction, is enabled by the shared motional modes of the ion chain, and we can perform a gate between any pair by on-demand driving of these modes to generate the desired entanglement. In particular, we use the IA beams, along with a global addressing beam (propagating counter to the IA beams), to perform the necessary motionally-sensitive red- and blue-detuned Raman transitions to generate each $\zz(\theta)$ gate pair~\cite{debnath2016, clark2021}. 

While the core two-qubit interaction in the system is about an equatorial rotation axis (XX-like)~\cite{sorensen2000}, we convert it to $\zz(\theta)$, via a phase-agnostic formulation~\cite{lee2005zz, tan2015, baldwin2020zz, campbell2023} that limits phase-dependent optical crosstalk~\cite{chow2024}. This conversion is achieved through the use of single-qubit counter-propagating ``wrapper'' gates which are also needed to combat the phase instabilities which arise from path length differences from mixing counter-propagating two-qubit gates and the standard co-propagating single-qubit gates~\cite{lee2005zz}. To realize the continuously parameterized entangling angle, $\theta$, the key degree of freedom exploited in this work, we adjust a Raman beam amplitude~\cite{shaffer2023} while keeping all $\zz(\theta)$ gates equivalent duration (250~$\mu$s). The gate has a Gaussian pulse shape and utilizes a constructive combination of radial motional modes (mode balancing) for frequency robustness when possible~\cite{ruzic2024}.

Typical single-qubit gate fidelities, $\R(\pi/2)$, are $\sim0.993$ while typical two-qubit gate fidelities, $\zz(\pi/2)$ are $\sim0.96 - 0.99$ depending on the pair. The virtual single-qubit phase gate, $Z(\theta)$, is assumed perfect. Estimated fidelities for each experimental run are provided in Sec.~\ref{sec:exp} and in Appendix~\ref{app:fid}. 

Empirically, we have observed that smaller entangling angle leads to improved two-qubit gate performance. Likewise, on other trapped-ion processors, a similar effect has been reported~\cite{moses2023}. More details on theoretical error scalings for QSCOUT's particular two-qubit gate configuration can be found in~\cite{ruzic2024}. Because of improved gate performance at smaller angles, here we aim to both develop and test compilation optimizations using \Superstaq/ to leverage this continuously parameterized two-qubit gateset within an all-to-all connected trapped-ion register.

\section{Compilation Optimizations} \label{Compilation}
\label{sec:comp}

Here we describe the basics of our circuit decomposition and optimization procedures for compiling arbitrary circuits for QSCOUT.
Because we expect (1) two-qubit gates to be the dominant source of error on QSCOUT, and (2) said error to scale roughly with the absolute rotation angle $|\theta| \le \pi/2$ of each $\zz(\theta)$ gate, the primary goals of our compiler optimizations are to minimize both the number and total angle of $\zz$ gates in the compiled circuit.

Our standard workflow for compiling circuits for QSCOUT comprises the following steps:
\begin{enumerate}
\item Merge adjacent two-qubit operations
\item Decompose the merged two-qubit unitaries into $\zz(\theta)$ operations and arbitrary single-qubit gates
\item Merge adjacent single-qubit operations
\item Decompose the merged single-qubit unitaries into $\R(\theta)$ operations and virtual phase gates
\end{enumerate}
Interspersed with these steps are various optimization passes to, for example, commute and merge single-qubit gates in order to minimize the number of physical operations required to execute the compiled circuit, or drop operations (such as diagonal gates at the start or end of the circuit) which have no logical effect.
Due to the register's all-to-all connectivity, no additional gates are needed to bridge missing qubit connectivity between two-qubit operations.

Single-qubit unitaries can always be implemented with a single physical pulse via their Euler decomposition,
\begin{equation}
  U = e^{i\gamma Z/2} e^{i\beta X/2} e^{i\alpha Z/2}
    = Z(\gamma+\alpha) R_\alpha(\beta) 
\end{equation}
where the leading term is implemented virtually.

In the following, we describe our baseline two-qubit gate implementation, and then four optimizations intended to improve performance over the baseline circuit construction. The first two reduce the overall total $\zz$ rotation angle. The next approach utilizes the all-to-all connectivity to reassign qubit roles based on gate performance, while the final optimization aims to reduce the overall number of two-qubit gates through circuit approximation.

\subsection{Baseline: Fixed Two-Qubit Gate}

As a baseline for comparisons, we restrict our gateset and decompose two-qubit unitaries to static $\zz(\pi/2)$ entangling gates --- ignoring (for now) the hardware's ability to implement arbitrary rotation angles.
A well-known result, Cartan's KAK decomposition, \cite{tucci2005} guarantees that any operation in $SU(4)$ can be decomposed into at most three such entangling pulses, interleaved with and surrounded by single-qubit operations on each qubit. 
As a result, the total $\zz$ rotation angle, $\Theta_U$, for each Haar-random $SU(4)$ operation that compose the quantum volume circuits considered is $\Theta_U = 3\pi/2$. 

\subsection{Continuously Parameterized Two-Qubit Gate}

We can decompose arbitrary two-qubit unitaries to QSCOUT's continuously-parameterized $\zz(\theta)$ operations by observing their KAK decompositions \cite{tucci2005}. Given an arbitrary unitary operation $U\in SU(4)$, we can always compute,
\begin{equation}
U = (C \otimes D) e^{ixXX/2} e^{iyYY/2} e^{izZZ/2} (A \otimes B)
\end{equation}
where $A, B, C, D$ are single-qubit unitaries and $\pi/4 \ge x \ge y \ge |z| \ge 0$ are the locally-invariant Weyl coordinates describing the entangling properties of the operation.
The $XX$ and $YY$ entangling terms can be converted to $\zz$ operations via conjugation by $R_{\pi/2}(\pi/2)^{\otimes 2}$ and $R_{0}(\pi/2)^{\otimes 2}$, respectively.
The total $\zz$ rotation angle $\Theta_U = x+y+\abs{z}\le3\pi/2$ is therefore upper-bounded by that of the baseline decomposition, with a mean (for Haar-random unitaries) of $3\pi/4$ --- half that of the baseline's static gate set.
The single-qubit rotations are then merged with the other single-qubit unitaries and decomposed into QSCOUT's gateset as before.

\subsection{Swap Mirroring}

As described in \cite{campbell2023}, we can further reduce the total $\zz$ cost by selectively inserting pairs of SWAP gates alongside existing two-qubit gates in the circuit. By performing one of the inserted SWAPs virtually (i.e. by relabelling qubits) and absorbing the second into the gate $U$ to be decomposed, we are left with a new gate $U'$ with Weyl coordinates $(\pi/4-x, \pi/4-y, z-\pi/4)$.
During the decomposition of merged two-qubit unitaries (step 2 of the compilation sequence outlined above), we therefore choose to insert SWAPs whenever $U'$ would require a smaller total $\zz$ angle than the original unitary $U$. This choice further reduces the maximum necessary $\Theta_U$ per selected unitary ($U$ or $U'$) to $3\pi/4$ and average to $2$ (or by just more than 15\%).
The net permutation incurred by the inserted virtual SWAPs is tracked in order to correctly map measurement outcomes on physical qubits to their original (circuit) positions.
QSCOUT's all-to-all connectivity ensures that the propagation of virtual SWAPs does not add any complexity in terms of routing to the compiled circuit.

\subsection{Gate Ranking}

In practice, interactions between different pairs of ions are not always equal. The fidelity of $\zz$ gates can vary between different qubit pairs due to the calibration of the drive lasers and geometry and interaction strength of the shared motional modes.
QSCOUT's all-to-all connectivity means we are free to arbitrarily assign circuit qubits to physical qubits (i.e. ions) on the processor.
We take further advantage of this connectivity in order to best utilize the highest-performing qubit pairs.
After inserting mirror SWAPs, we sum the total rotation angle of $\zz$ gates on each pair of circuit qubits in the decomposed circuit. We then choose a physical qubit assignment by minimizing $\sum_{ij}\epsilon_{ij}|\theta_{ij}|$, where $\theta_{ij}$ is the $\zz$ rotation angle that would occur on physical qubits $(i,j)$ in the mapped circuit, and $\epsilon_{ij}$ is the expected rate of error for $\zz$ gates on those qubits.
For the small circuits considered here we employ a simple brute-force approach to optimize qubit assignment; however we find that a more scalable greedy approach can perform nearly as well.

\subsection{Circuit Approximation}

Sometimes, the compilation processes above results in gates with very small rotation angles. Though we expect that gate error to increases with rotation angle, we also expect a finite amount of error due to doing the gate at all (both due to the implementation of the gate itself, and the extended runtime due to the constant duration of $\zz$ operations). For a gate that is sufficiently well-approximated by an identity operation, these errors can outweigh those which would be incurred from eliding the gate altogether. To this end, the compilation stack includes a variable tolerance parameter fixing the minimum rotation angle $\theta_{min}$ below which gates will be dropped.

Approximate decomposition provides further opportunity for gate elimination when used in conjunction with mirror swapping. In this case, if a gate $\zz(\theta)$ is sufficiently well-approximated by a $\pi/2$ rotation (i.e. $|\theta - \pi/2| \le \theta_{min}$), the insertion of a mirror swap will mean we can drop that gate entirely. We therefore make the mirror swapping insertion pass aware of the approximation tolerance parameter, and always insert (do not insert) a mirror swap if the resulting decomposition will require fewer (more) entangling operations.
A similar technique was employed in~\cite{cross2019}, in which mirror swaps were used to reduce the number of CNOT gates required for approximating two-qubit unitaries in quantum volume circuits.

\section{Evaluation Approaches for Compilations with Random Circuits} \label{sec:QV}

\subsection{Circuit Choice}
To examine the impact of these various compilation optimizations harnessing the continuously parameterized gateset and full connectivity, we generate randomized circuits across all qubits in the register. These types of circuits are used in the prevalent quantum volume benchmarking approach~\cite{cross2019, baldwin2022qv, pelofske2022}. In particular, these randomized circuits are a square measure consisting of $n$ ions at a depth $d = n$ layers. Each layer within the circuit consists of Haar-random SU(4) operations applied to randomly selected pairs of qubits. For each layer, a permutation is performed such that new pairings are available for subsequent layers. An important point is that in some cases (especially for the qubit register size demonstrated here, $n=4$ or 5), the permutation might select the same pairing(s) from the previous layer in which case the two SU(4) operations can be merged. As described in Section~\ref{sec:comp}, this approach is used for all circuits.

\subsection{Analysis Approaches}
\label{sec:analysis_approaches}

Here we introduce three analysis approaches to quantify improvements realized by the various compilation techniques. The first two are variants on the typical heavy output analysis, one we call the ``unaware heavy output probability,'' $\hunaware$, as well as the standard approach, which we term the ``aware heavy output probability,'' $\haware$. The final analysis approach is the Hellinger infidelity, $\infidelity$, used to measure the closeness of the empirical probability distribution to the ideal distribution. In addition to quantifying performance improvements, these techniques also help to elucidate error channels in our system. We now describe each in detail below.

The first metric, the unaware heavy output, $\hunaware$, is the sum of all output probabilities greater than the median of the resulting probability distribution. This rudimentary version is only susceptible to incoherent/stochastic errors (dephasing, depolarizing, etc.), and not to coherent errors because it is ignorant of the ideal outputs and only examines the empirical outputs, $M$, of a given circuit to assess its heaviness. For all possible observable $2^n$ qubit states (bit strings), $q \in \{0, 1\}^n$, we find the states $q_M$: 
\begin{equation}
q_M = \{q \in \{0, 1\}^n \textrm{ such that } M(q) >  \tilde{M}\}
\end{equation}
where $\tilde{M}$ is the median of the set of (measured) output probabilities, $M$. The unaware heavy output, $\hunaware$ is thus:
\begin{equation}
\hunaware = \sum_{q \in q_M} M(q)
\end{equation}
Because the median $\tilde{M}$ is computed from the empirical output distribution, measurement of $\hunaware$ requires the number of measurements per circuit to be large relative to $2^n$.

The second metric is the standard approach to quantum volume benchmarking. The aware heavy output, $\haware$ is the sum of those empirical outputs whose \emph{corresponding ideal outputs (simulated)} lie above the ideal median~\cite{cross2019, baldwin2022qv}, thus penalizing both coherent errors and stochastic errors. With this measure, we find the states $q_S$ from the ideal simulated output distribution, $S$, such that:
\begin{equation}
q_S = \{q \in \{0, 1\}^n \textrm{ such that } S(q) >  \tilde{S}\}
\end{equation}
where $\tilde{S}$ is the median of the $S$ ideal set. We take only those qubit states whose \emph{ideal} output is classified as heavy, and then sum the corresponding empirical outputs:
\begin{equation}
\haware = \sum_{q \in q_S} M(q)
\end{equation}
Because the median $\tilde{S}$ is computed from the ideal distribution rather than the empirical output, measurements of $\haware$ do not depend on having a high number of shots per circuit (provided the number of random circuits is sufficiently high) --- projection noise instead depends on the number of measurements taken across all random circuits. This can make this measurement significantly less costly experimentally, especially for larger numbers of qubits. However, we find that computing $\hunaware$ alongside $\haware$ can provide us with some unique insights into the dominant errors impacting the system, in exchange for this extra cost.

In the limit of large $n$, we expect both heavy output probabilities to approach $\mathbb{E}[\hunaware] = \mathbb{E}[\haware] = (1 + \ln 2)/2\approx0.847$ for Haar-random pure quantum states. For smaller $n$ this value needs to be reduced slightly \cite{baldwin2022qv} --- for example, we find empirically that $\mathbb{E}[\hunaware]=\mathbb{E}[\haware]\approx0.839$ for Haar-random states at $n=5$, and $\mathbb{E}[\hunaware]=\mathbb{E}[\haware]\approx0.831$ at $n=4$. Approaching a fully mixed state, we expect the outputs to approach a uniform distribution and therefore the probability of measuring a heavy output to approach $1/2$ (independent of $n$).

For $\hunaware$ this floor increases slightly when the number of samples approaches $2^n$, as the expected median $\mathbb{E}[\tilde{M}]$ is biased below the mean $\mathbb{E}[M] = 2^{-n}$ due to projection noise.
For example, we find $\mathbb{E}[\hunaware]\approx0.554$ for 200 samples of a uniform distribution with $n=4$.
Because $\haware$ uses the ideal distribution to compute $\tilde{S}$, it is not impacted by sample size in the same way.

We also note that the $\hunaware$ and $\haware$ measures are typically understood in terms of their mean corresponding to a large number of random circuits. From this, a quantum volume of $2^n$ is certified when both the $\haware$ mean and its $2\sigma$ confidence interval lower bound lie above the 2/3 threshold for $n$ qubits at a depth of $d = n$ layers. This 2/3 threshold is inspired by theoretical studies of quantum advantage and speedup~\cite{aaronson2016}. Traditionally, the confidence interval corresponds to a Wald interval~\cite{cross2019}, but more recently, bootstrapping methods are able to yield tighter confidence intervals resulting in fewer circuits needed to certify the quantum volume~\cite{baldwin2022qv}. In this work, however, we will use a more conservative measure than the Wald interval, the Wilson score interval, to determine $2\sigma$ intervals. The Wilson score interval is asymmetric and avoids the potential overshoot issues associated with the Wald interval especially for smaller sample sizes~\cite{wilson1927}. At significantly large numbers of circuits sampled, and with means sufficiently far from the limits, the Wald and Wilson score intervals converge.  

The final metric we use is Hellinger infidelity. Hellinger fidelity is a measure similar to classical fidelity but is derived from the Hellinger distance, $d_{H}$, a measure of the distance between two probability distributions. Here the two distributions are the ideal simulated distribution $S$, and the experimentally observed distribution $M$. Thus, $d_{H}$ is defined:
\begin{equation}
d_H(S,M) = \frac{1}{\sqrt{2}} \sqrt{\sum_{q \in \{0, 1\}^n} (\sqrt{S(q)} - \sqrt{M(q)})^2}
\end{equation}
The Hellinger fidelity, $\mathcal{F}_{H}$ is then:
\begin{equation}
\mathcal{F}_{H} = (1 - d_{H}^2)^2
\end{equation}
And the Hellinger infidelity, $\infidelity$ is simply:
\begin{equation}
\infidelity = 1 - \mathcal{F}_{H}
\end{equation}

Like the unaware heavy output probability, measurements of Hellinger infidelity are impacted by shot noise. For example, with 200 shots we expect shot noise to increase infidelities by about 0.02 for 4-qubit QV circuits and 0.05 for 5-qubit QV circuits. Thus, in addition to a sufficient number of distinct circuits, a sufficient number of shots per circuit should be used.

All three measures help to give a picture of the underlying error sources of the continuously parameterized two-qubit gate. For each compilation optimization, we generally find improvements in all measures, but some demonstrate a stronger improvement than other measures. It is important to note each measure has its limitations due to sampling or convolution of differing effects, and utilizing all three provides a more holistic picture of the comparison.
They also diverge in terms of experimental overhead and scalability, with $\haware$ requiring significantly fewer shots than either $\hunaware$ or $\infidelity$.

\subsection{Distinguishing errors}

The unaware heavy output probability $\hunaware$ is totally independent of the ideal circuit outputs. Instead, it characterizes the degree to which measurement outcomes align with the expectations from random circuit sampling, i.e. how well they converge upon the expected Porter-Thomas distribution of output probabilities.
In most cases a circuit with coherent (unitary) errors still resembles a random circuit---but not the circuit we are attempting. Because the calculation of $\hunaware$ is agnostic toward the ideal circuits, it is largely insensitive to these coherent errors so long as the collection of faulty circuits is still approximately Haar-random.
While there are cases for which Haar-randomness is not preserved (take for example the limiting case of systematic 100\% under-rotation of all gates, i.e. no gates are performed at all), we find that in practice this divergence is negligible for the types and magnitudes of coherent errors that could feasibly be present in our experiments.
Stochastic errors on the other hand are not information-preserving, and so are readily observed as changes in $\hunaware$.

On the other hand, the aware heavy output probability $\haware$ \emph{does} take into account the ideal output distribution of the circuit, and therefore is expected to reflect both stochastic and coherent errors.
The comparison between these values provides us with a simple heuristic for distinguishing between certain classes of errors: with no errors, we expect the two metrics to be identical; for stochastic noise we expect a similar shift in heavy outputs from both metrics; and for coherent errors we expect a divergence between the two.

\begin{figure}[h]
    \centering
    \includegraphics[width=0.45\textwidth]{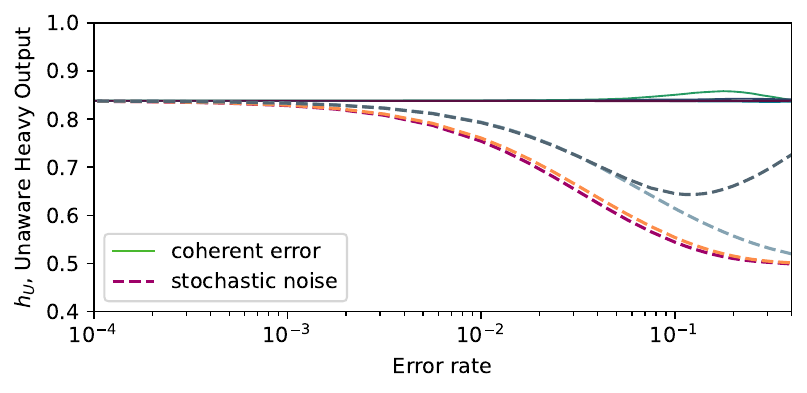}
    \caption{Simulated unaware heavy output probabilities ($\hunaware$) of 400 4-qubit QV circuits as a function of error rate for a variety of coherent (solid lines of varying color) and stochastic (dashed lines of varying color) error models. 
    Note that only the stochastic errors result in a meaningful reduction in heavy outputs.
    }
    \label{fig:sim:hunaware}
\end{figure}

\begin{figure}[h]
    \centering
    \includegraphics[width=0.45\textwidth]{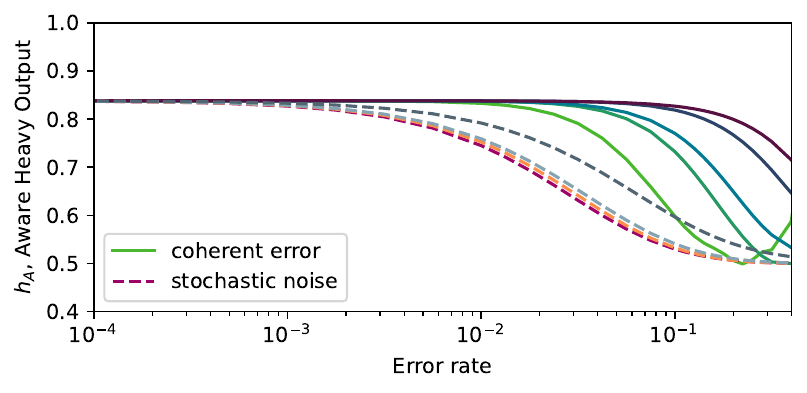}
    \caption{Simulated aware heavy output probabilities ($\haware$) for 400 QV 4-qubit circuits as a function of error rate for a variety of coherent (solid lines of varying color) and stochastic (dashed lines of varying color) error models.
    }
    \label{fig:sim:haware}
\end{figure}

We validate this behavior empirically by simulating 400 compiled 4-qubit QV circuits subject to a variety of coherent error sources (e.g. relative over-rotations of one- or two-qubit gates and angle-dependent phase errors on single-qubit gates) and stochastic noise channels (e.g. depolarization affecting just one- or two-qubit gates, time-dependent heating, and dephasing). The resulting unaware and aware heavy outputs are presented as a function of error rate in \cref{fig:sim:hunaware,fig:sim:haware}, respectively. As expected, we see a clear dependence of $\hunaware$ on error rate for all of the stochastic error channels (dashed lines). This dependence disappears for the coherent channels (solid lines), which show almost no variation except at the highest rates (which are well beyond what we could reasonably expect to see on the hardware). For the same set of circuits, we see that $\haware$ has a clear dependence on error rate for both the stochastic and coherent channels.
In \cref{sec:error_channels} we will use this analysis to estimate the amount of stochastic noise in our experiments, and observe how it is affected by the various compiler optimizations described in \cref{Compilation}.

\section{Empirical Comparisons of Compilation Approaches}
\label{sec:exp}
With these various evaluations, we now turn our attention to comparing these compilation optimizations on hardware, the trapped-ion QSCOUT system described in Section~\ref{sec:qscout}.

\subsection{Continuously Parameterized Two-Qubit Gate}
We begin by comparing a set of 200 randomly selected quantum volume circuits, generated via \qiskit/'s quantum volume library~\cite{qiskit2024}, on a 4-qubit register with a depth of 4 layers. These 200 circuits are first compiled via \Superstaq/ with only fixed entangling gates, $\zz(\pi/2)$ as described in Section~\ref{sec:comp} to act as our baseline. These same circuits are then compiled using the continuously parameterized entangling gate, $\zz(\theta)$. During the experimental run, the circuits from the different sets (along with the swap mirrored circuits described in the Section~\ref{sec:exp_swap_mirror}) are interleaved such that they are all susceptible to the same noise and drift during the course of the experimental run. In particular, each version of the circuit is run 200 times before moving onto the next variation of that circuit. After all variations of a particular circuit are run, the next distinct circuit and its different compiler variations are run. The resulting qubit state output probabilities are then analyzed with the methods described in Section \ref{sec:QV}. 

\begin{figure}[h]
    \centering
    \includegraphics[width=0.45\textwidth]{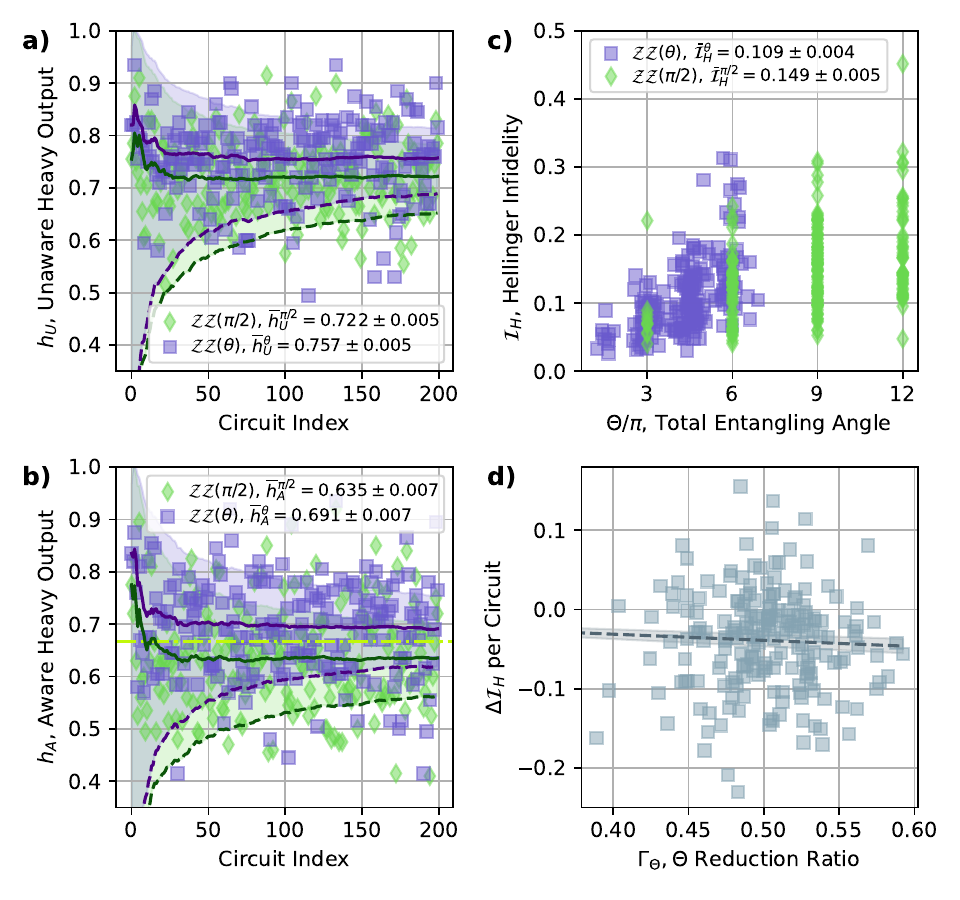}
    \caption{Comparison of 200 quantum volume circuits on a 4-qubit register compiled with either fixed entangling gates, $\zz(\pi/2)$ (green diamonds), or continuously parameterized entangling gates, $\zz(\theta)$ (purple squares). Analysis with a) unaware heavy outputs versus run-time circuit ordering, b) aware heavy outputs versus run-time circuit ordering, c) Hellinger infidelity vs. total entangling angle, $\Theta$, and d) per circuit change in Hellinger infidelity versus overall $\Theta$ reduction ratio, $\Gamma_{\Theta}$. In a) and b), each running mean is presented via a solid line, while $2\sigma$ Wilson score intervals are presented with shading in between the bounds with the lower bound presented via a dashed line. In d), a trend line is fit to this data and the data in Fig.~\ref{fig:mirrorSWAP}d with a 2$\sigma$ error envelope.}\label{fig:cont}
\end{figure}

In Fig.~\ref{fig:cont}a, we see a significant increase in $\hunaware$ for the circuits compiled with $\zz(\theta)$ relative to those compiled with $\zz(\pi/2)$. Likewise, in Fig.~\ref{fig:cont}b, while we see the overall heaviness is reduced in both cases due to additional coherent error, the relative increase in $\haware$ of the $\zz(\theta)$ circuits over the $\zz(\pi/2)$ circuits is similar. This is particularly striking as it transforms a series of circuits whose $\haware$ mean probability lies below the quantum volume threshold of 2/3 to a series of circuits whose $\haware$ mean probability surpasses the threshold (albeit without a sufficient number of circuits for the $2\sigma$ confidence intervals to surpass the threshold). 

To avoid convolving additional randomness from the varied natural heaviness of circuits, we instead use the $\infidelity$ to elucidate potential trends in the data. In Fig.~\ref{fig:cont}c, we plot $\infidelity$ versus the overall entangling angle (summed over all gates and qubits in the circuit) i.e. $\Theta= \sum_{i}\abs{\theta_i}$ where $\theta_i$ is the entangling angle for each $\zz(\theta)$ gate within the circuit. Here, we see a clear trend of increasing infidelity with increasing $\Theta$. There is also distinct clustering in the data which is reflective of the role that SU(4) merges play on the circuit length. There are four sets of clusters along $\Theta$ in each dataset (note, that the clusters of the $\zz(\theta)$ compilation are more apparent in Fig.~\ref{fig:mirrorSWAP}). Each cluster within one compilation corresponds to a cluster in the other e.g. the $\Theta=12\pi$ cluster of the $\zz(\pi/2)$ compilation corresponds to the same sets of circuits in $\Theta~6\pi$ cluster of the $\zz(\theta)$ compilation. We see each $\zz(\theta)$ cluster occupies a lower Hellinger infidelity range than its corresponding cluster for $\zz(\pi/2)$.

Next, we examine $\infidelity$ on a per circuit basis in Fig.~\ref{fig:cont}d. Here, each circuit is sorted by how much the total $\Theta$ has been reduced via a reduction ratio, $\Gamma_{\Theta}$ defined as:
\begin{equation}\label{eq:reduction_ratio}
\Gamma_{\Theta} = \frac{\Theta_{i} - \Theta_{j}}{\Theta_{i}}
\end{equation}
in which $i$, $j$ refer to the two compilation optimizations being used here, the baseline $\zz(\pi/2)$ vs. $\zz(\theta)$ compilations respectively. We see the entangling angle is reduced by $\Gamma_{\Theta}\sim0.5$, and in general, we see a reduction in the per circuit $\infidelity$ as a result. A trend line is presented in which this $\Delta\infidelity$ dataset and the $\Delta\infidelity$ dataset from Fig.~\ref{fig:mirrorSWAP}d are both fit together, indicating a slight trend toward decreasing $\infidelity$ with increasing $\Gamma_{\Theta}$. 
\label{sec:exp_cont}

\subsection{Swap Mirroring}
\begin{figure}[h]
    \centering
    \includegraphics[width=0.45\textwidth]{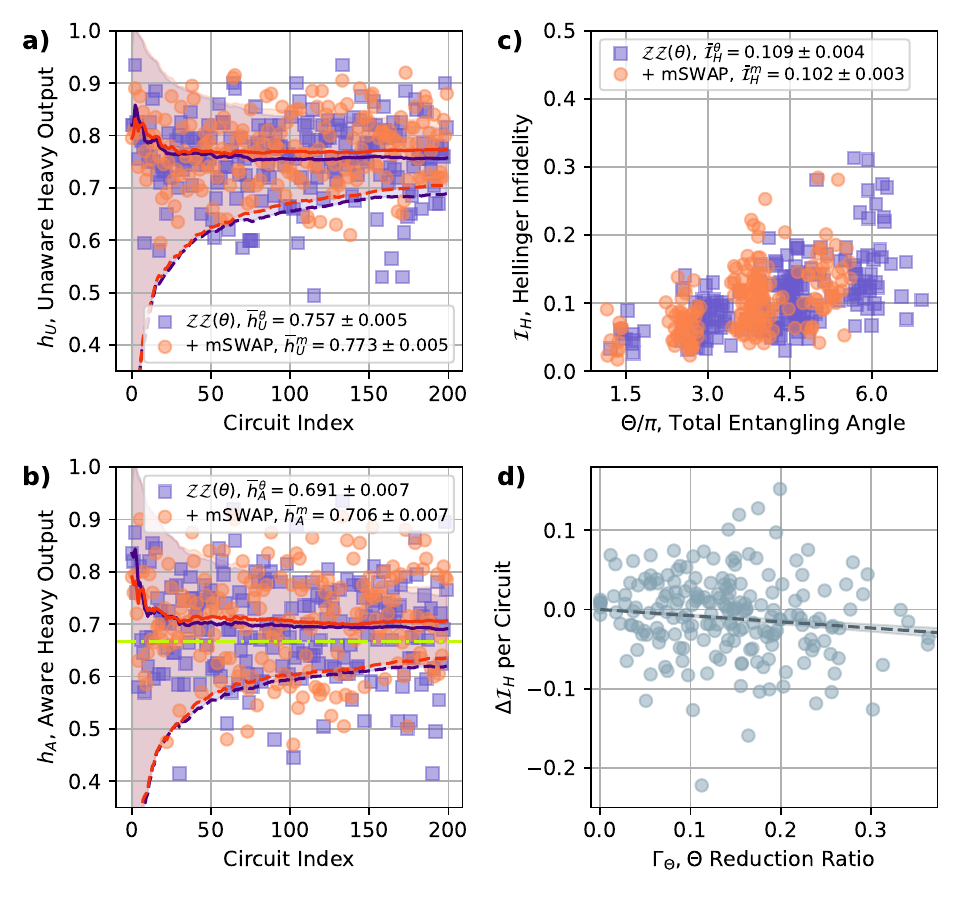}
    \caption{Comparison of 200 quantum volume circuits on a 4-qubit register compiled with continuously parameterized entangling gates, $\zz(\theta)$ with swap mirroring (orange circles, denoted ``+mSWAP'' in this and subsequent figures for conciseness) and without (purple squares). Analysis with a) unaware heavy outputs versus run-time circuit ordering, b) aware heavy outputs versus run-time circuit ordering, c) Hellinger infidelity vs. total entangling angle, $\Theta$, and d) per circuit change in Hellinger infidelity versus overall $\Theta$ reduction ratio, $\Gamma_{\Theta}$. In a) and b), each running mean is presented via a solid line, while $2\sigma$ Wilson score intervals are presented with shading in between the bounds with the lower bound presented via a dashed line. In d), a trend line is fit to this data and the data in Fig.~\ref{fig:cont}d with a 2$\sigma$ error envelope. Note, Figs.~\ref{fig:cont} and \ref{fig:mirrorSWAP} are the same 200 circuits, in which all three compilation choices were taken interleaved with one another.}
    \label{fig:mirrorSWAP}
\end{figure}

With those same 200 circuits described above, the next variation we test continues to compile the circuits with $\zz(\theta)$ but with the added feature of swap mirroring described in Sec.~\ref{sec:comp}. This experiment was interleaved (as described above) with the ones in Fig.~\ref{fig:cont}. For simplicity, we present only the comparison of the $\zz(\theta)$ compilations with and without swap mirroring in Fig.~\ref{fig:mirrorSWAP}.
\label{sec:exp_swap_mirror}

The improvements for this approach are more subtle than those seen in switching from a fixed entangling gate to the continuously parameterized entangler. Both the $\hunaware$ and $\haware$ measures show distinct but smaller improvements in performance in Figs.~\ref{fig:mirrorSWAP}a and \ref{fig:mirrorSWAP}b as compared to Figs.~\ref{fig:cont}a and \ref{fig:cont}b. 

These improvements are also seen in the $\infidelity$ measures as well presented in Fig.~\ref{fig:mirrorSWAP}c. We once again see the four distinct clusters from the differing degrees of SU(4) merges, and the swap mirrored circuits all inhabit clusters at lower entangling angle than those without swap mirroring. Likewise, when the per-circuit $\Theta$ reduction, $\Gamma_{\Theta}$ is plotted for these two compilation approaches, we once again see a reduction in the $\infidelity$ (Fig.~\ref{fig:mirrorSWAP}d) along with the same trend for decreasing infidelity with increasing $\Gamma_{\Theta}$ as well. 

These more marginal improvements in both heavy output measures and $\infidelity$ are to be expected as the overall reduction of total entangling angle is $\Gamma_{\Theta}\sim0.15$ on average across all of the circuits. However, we have previously demonstrated swap mirroring on a class of circuits in which the choice to swap mirror always reduces the entangling angle and saw more drastic improvements in performance on the QSCOUT system as the overall average $\Gamma_{\Theta}\sim0.61$ for that collection of circuits~\cite{campbell2023}.

\subsection{Gate Ranking}
The next approach we investigate relies on knowledge of the machine's performance on a gate-by-gate basis. In particular, the compiler will attempt to reduce the total entangling angle on the lesser performing two-qubit gates. For these demonstrations, we intentionally depress the performance of certain gates in order to highlight this compilation approach. Here, we will compile all circuits with $\zz(\theta)$ entanglers and swap mirroring enabled. 
\label{sec:exp_gate_ranking}

\subsubsection{``Bad Pairs''}
\label{sec:bad_pairs}

\begin{figure}[h]
    \centering
    \includegraphics[width=0.45\textwidth]{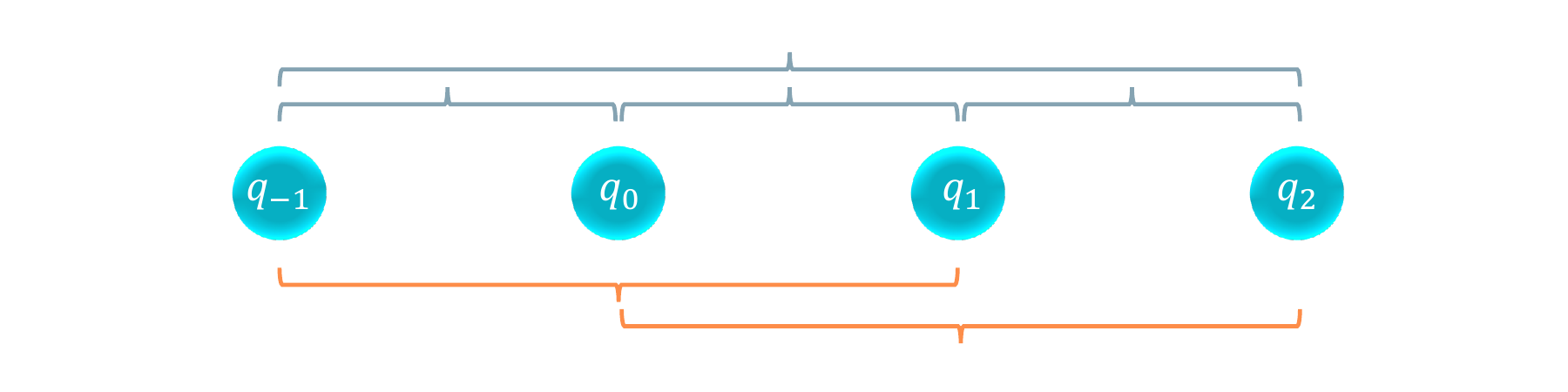}
    \caption{Four-ion chain with all-to-all connectivity. Nearest neighbor interactions and outer interaction (in grey brackets) are $\sim0.99$ estimated fidelity. Next-nearest-neighbor interactions (in orange brackets) intentionally depressed to $\sim0.95$ estimated fidelity.}
    \label{fig:badpairs_cartoon}
\end{figure}

\begin{table}[h]    
    \begin{tabular}{|c|c|c|}
                \hline
            Gate Pair & Connectivity & Est. State Fidelity \\
        \hline
        \{-1, 0\} & nearest neighbor & $0.987_{-0.005}^{+0.004}$ \\
        \hline
        \{0, 1\} & nearest neighbor & $0.987_{-0.005}^{+0.004}$ \\
        \hline
        \{1, 2\} & nearest neighbor & $0.989_{-0.005}^{+0.004}$ \\
        \hline
        \rowcolor{orange!20}
        \{-1, 1\} & next-nearest neighbor & $0.954_{-0.007}^{+0.007}$ \\
        \hline
        \rowcolor{orange!20}
        \{0, 2\} & next-nearest neighbor & $0.952_{-0.007}^{+0.007}$ \\
        \hline
        \{-1, 2\} & outer & $0.988_{-0.005}^{+0.004}$ \\
        \hline
    \end{tabular}
\caption{Fidelity estimates of two-qubit gate pairs in a four-ion chain.  Qubit name assignments are in Fig.~\ref{fig:badpairs_cartoon}. Performance of gate pairs \{-1, 1\} and \{0, 2\}, highlighted in orange, has been intentionally depressed through a mismatch of amplitudes of the red- and blue-detuned Raman transitions during the $\zz(\theta)$ interaction. To \emph{estimate} the fidelity of the fully entangling interaction efficiently prior to runtime, we repurpose our AC Stark cancellation calibration consisting of two successive bare $MS(\pi/2)$ gate interactions (no ``wrapper'' gates to create the $\zz(\pi/2)$ interaction) while varying the \emph{in-situ} Z gate to cancel the AC Stark shift. We determine the greatest probability of $\ket{11}$ through a maximum likelihood estimation fit and extract its $2\sigma$ confidence intervals. To then estimate the fidelity of a maximally entangled state generated by a single gate, the square root of the $\ket{11}$ probability is used, while the uncertainties are derived from its $2\sigma$ confidence intervals.}\label{tab:bad_pairs} 
\end{table}

The first demonstration uses a four-qubit register, where we find the pairs between next-nearest neighbors are typically lower fidelity operations, see Fig.~\ref{fig:badpairs_cartoon}. These operations also incur more optical crosstalk on the skipped qubit, so they are ideal for compiling around. Here, to better emphasize the technique, we mismatch the red- and blue-detuned Raman amplitudes to depress the next-nearest neighbor gates beyond their natural fidelities. To be exact, we decrease the blue-detuned Raman amplitude to $80\%$ of the optimal performance. This incurs additional AC Stark shift on the participating qubits as the detuned carrier Stark shifts from red- and blue-detuned drives no longer cancel~\cite{lee2016starkshifts}.  This additional physical AC Stark shift is cancelled by a virtual \emph{in-situ} Z gate, but the resulting additional dephasing noise from the stronger AC Stark shift remains.

In Table~\ref{tab:bad_pairs}, we list the measured state fidelity estimates determined by measuring the $\ket{11}$ state after two successive $MS(\pi/2)$ gates (i.e. $\zz(\pi/2)$ without the ``wrapper'' single-qubit gates to transform its basis). As this is a rough measure and used for relative comparisons as opposed to absolutes, no SPAM correction is implemented. These values act as an upper bound on the true fidelity of the operation, and we see, after our intentional miscalibration, the next-nearest neighbor pairs are limited to an upper bound of $\sim0.953$ for their fidelity.

\begin{figure}[b]
    \centering
    \includegraphics[width=0.45\textwidth]{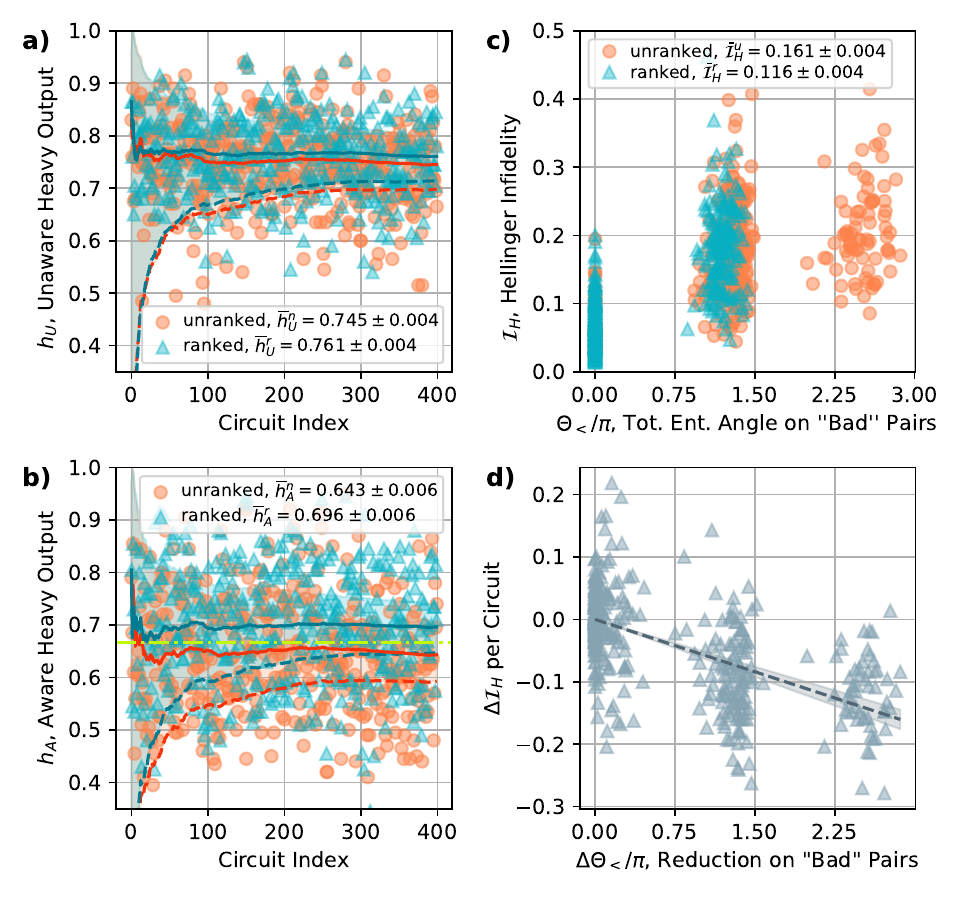}
    \caption{Comparison of 400 quantum volume circuits on a 4-qubit register (compiled with continuously parameterized entangling gates, $\zz(\theta)$ with swap mirroring by default). with compilation that aims to put the largest entangling angle on the best performing gates (teal triangles) versus no knowledge of gate performance (orange circles). Here, the two orthogonal (and next-nearest neighbor) gate pairs were intentionally made worse. Analysis with a) unaware heavy outputs versus run-time circuit ordering, b) aware heavy outputs versus run-time circuit ordering, c) Hellinger infidelity vs. total entangling angle on the poorer performing gates, $\Theta_{<}$, and d) per circuit change in Hellinger infidelity versus overall $\Theta_{<}$ reduction on the poorer performing gates. In a) and b), each running mean is presented via a solid line, while $2\sigma$ Wilson score intervals are presented with shading in between the bounds with the lower bound presented via a dashed line. In d), a trend line is presented with a 2$\sigma$ error envelope.}
    \label{fig:badpairs}
\end{figure}

We compile 400 randomly selected quantum volume circuits with knowledge of the poorer performing gates, referred to as ``ranked'' circuits, as well as without knowledge, referred to as ``unranked.'' We find of the 400 circuits, $\sim2/3$ of the circuits are able to take advantage of this compilation approach, and within that subset, more than half are able to remove participation of those gate pairs completely, see Appendix~\ref{app:stats}.

The circuits are once again performed interleaved with one another at 200 shots, and we find that the there is improvement in all three analysis measures in Fig. \ref{fig:badpairs}. Interestingly, we see a smaller gap between the $\hunaware$ measures than in the $\haware$ measures, suggesting we introduced substantial coherent error on our $\zz(\theta)$ gates on the next-nearest neighbor pairs, Figs.~\ref{fig:badpairs}a and \ref{fig:badpairs}b. These error channels are explored further in Section~\ref{sec:error_channels}. In terms of quantum volume, when using the ``unranked'' circuits, which are agnostic to gate performance, the $\haware$ mean remains below the quantum volume threshold of 2/3; however, when compiling with knowledge of the ``bad pairs,'' the $\haware$ mean now lies above that threshold (once again, without a sufficient number of circuits for the $2\sigma$ confidence intervals to cross the threshold thus certifying quantum volume).

In (Fig.~\ref{fig:badpairs}c), the $\infidelity$ measure shows a noisy but distinct worsening trend for increasing entangling angle on the lesser-performing gate pairs, i.e. $\Theta_{<} = \sum_{i}\theta_{i,p}$ for all pairs $p$ that are lesser-performing. When examined on a per circuit basis, and sorting the circuits based on their how much total entangling angle has been removed from the ``bad'' pairs, $\Delta\Theta_{<}$, and reassigned to better performing pairs, i.e. ``offloaded,'' we see a clear reduction trend in the $\infidelity$ for increased offloading from those next-nearest neighbor pairs (Fig.~\ref{fig:badpairs}d). Here, however, the clustering is less reflective of SU(4) merges and instead reflects different scenarios for how much entangling angle is offloaded from the poorer performing gates and is described further in Appendix~\ref{app:stats}.
    
\subsubsection{``Bad Qubit''}
\label{sec:bad_qubit}

Another scenario we evaluate is the case where one qubit's $\zz$ gate pairs all perform worse than the other pairs available in the chain. Here, we increase the qubit register to $n=5$ (and similarly, the circuit depth, $d=5$) to better demonstrate this scenario. In particular, quantum volume circuits differ substantially between even and odd-numbered registers. As the circuit construction relies upon pairing qubits within an SU(4) operation, odd registers will always have one qubit that does not participate within any given layer~\cite{cross2019, baldwin2022qv}. Therefore, for any given circuit, this gate ranking scenario essentially recasts the least-participatory qubit to be the worst-performing one.

\begin{figure}[h]
    \centering
    \includegraphics[width=0.45\textwidth]{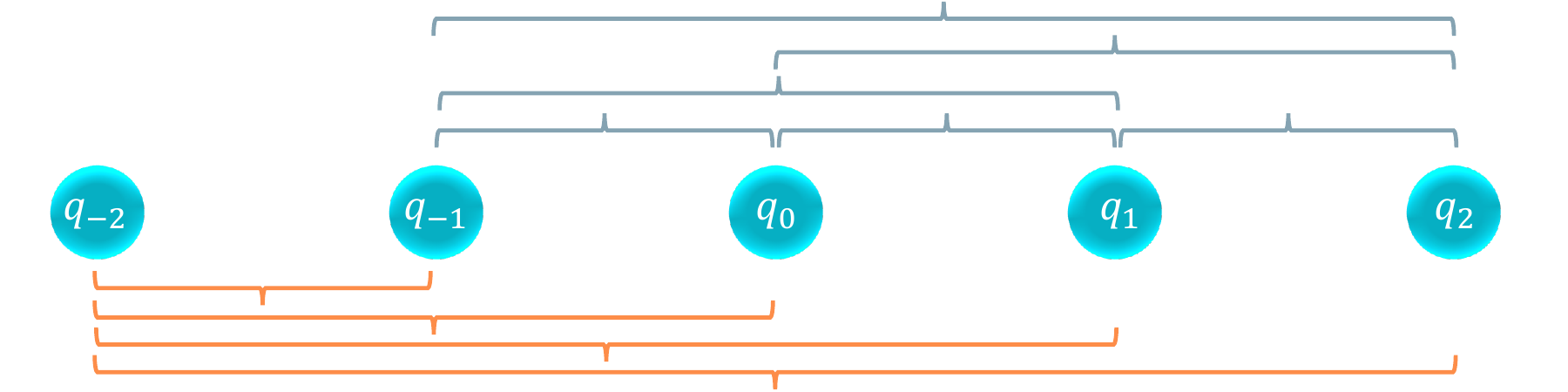}
    \caption{Five-ion chain with all-to-all connectivity. All interactions with one of the outer qubits, $q_{-2}$, are intentionally depressed in estimated fidelity to $\sim0.89-0.92$ (in orange brackets), while all other interactions have fidelity estimates $\sim0.95-0.97$ (in grey brackets).}
    \label{fig:badqubit_cartoon}
\end{figure}

Here, the left-most qubit in the chain, $q_{-2}$, is deemed to be the ``bad qubit,'' (Fig. $\ref{fig:badqubit_cartoon}$). Here, we intentionally inject error into all of its corresponding gate pairs to ensure all its pairs' fidelity estimates are below every other $\zz$ gate pairs via the same mismatch of blue- and red-detuned amplitudes (at a ratio of 80\%) that was done for the ``bad pairs'' scenario. Fidelity estimates are provided in Table~\ref{tab:bad_qubit}. 

\begin{table}    
    \begin{tabular}{|c|c|c|}
         \hline
                Gate Pair & Connectivity & Est. State Fidelity \\
            \hline
            \rowcolor{orange!20}
            \{-2, -1\} & nearest neighbor & $0.897_{-0.010}^{+0.010}$ \\
            \hline
            \{-1, 0\} & nearest neighbor & $0.965_{-0.008}^{+0.007}$ \\
            \hline
            \{0, 1\} & nearest neighbor & $0.961_{-0.008}^{+0.007}$ \\
            \hline
            \{1, 2\} & nearest neighbor & $0.968_{-0.008}^{+0.007}$ \\
            \hline
            \rowcolor{orange!20}
            \{-2, 0\} & next-nearest neighbor & $0.916_{-0.010}^{+0.009}$ \\
            \hline
            \{-1, 1\} & next-nearest neighbor & $0.961_{-0.008}^{+0.007}$ \\
            \hline
            \{0, 2\} & next-nearest neighbor & $0.959_{-0.008}^{+0.008}$ \\
            \hline
            \rowcolor{orange!20}
            \{-2, 1\} & third-nearest neighbor & $0.922_{-0.009}^{+0.009}$ \\
            \hline
            \{-1, 2\} & third-nearest neighbor & $0.947_{-0.009}^{+0.008}$ \\
            \hline
            \rowcolor{orange!20}
            \{-2, 2\} & outer & $0.890_{-0.010}^{+0.010}$ \\
            \hline

    \end{tabular}
\caption{Fidelity estimates of two-qubit gate pairs in a five-ion chain.  Qubit name assignments are found in Fig.~\ref{fig:badqubit_cartoon}. The performance of all gate pairs involving $q_{-2}$, highlighted in orange, have been intentionally depressed through a mismatch of amplitudes of the red- and blue-detuned Raman amplitudes during the $\zz$ interaction. As described in the caption of Table~\ref{tab:bad_pairs}, fidelity estimates are determined via measurements taken just prior to runtime.}\label{tab:bad_qubit} 
\end{table}

\begin{figure}[h]
    \centering
    \includegraphics[width=0.45\textwidth]{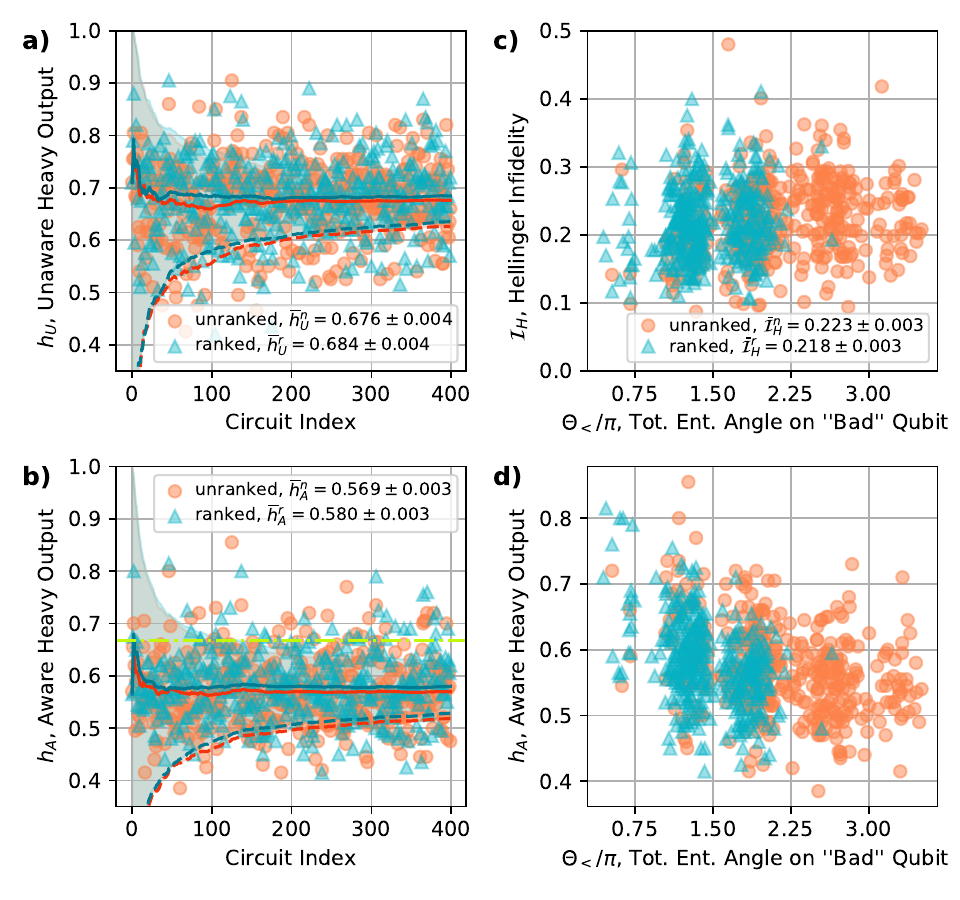}
    \caption{Comparison of 400 quantum volume circuits on a 4-qubit register (compiled with continuously parameterized entangling gates, $\zz(\theta)$ with swap mirroring by default). with compilation that aims to put the largest entangling angle on the best performing gates (teal triangles) versus no knowledge of gate performance (orange circles). Here, pairs involving $q_{-2}$ were intentionally made worse. Analysis with a) unaware heavy outputs versus run-time circuit ordering, b) aware heavy outputs versus run-time circuit ordering, c) Hellinger infidelity vs. total entangling angle on the poorer performing gates, $\Theta_{<}$, and d) aware heavy outputs versus overall $\Theta_{<}$ reduction. In a) and b), each running mean is presented via a solid line, while $2\sigma$ Wilson score intervals are presented with shading in between the bounds with the lower bound presented via a dashed line.}
    \label{fig:badqubit}
\end{figure}

To examine this scenario, we generate 400 randomly selected circuits, once again performed at 200 shots, and interleaved.  Here we find small yet distinct improvements in the $\hunaware$ and $\haware$ measures (Figs. \ref{fig:badqubit}a and \ref{fig:badqubit}b). We note these improvements are more subtle than the case of the ``bad pairs'' scenario, likely due to less overall average offloading of entangling angle for this scenario. We also see only a marginal improvement in the $\infidelity$ measure (Fig. \ref{fig:badqubit}c). The marginal change in the $\infidelity$ is partially due to sample size susceptibility of this analysis approach (described in Sec.~\ref{sec:analysis_approaches}) as a five-qubit chain has twice as many qubit bitstrings as a four-qubit chain, leading the measure to be subsumed by more shot noise (for the same number of shots) and thus an increased floor. Instead, to examine the effects on a per circuit basis, we look at the $\haware$ measure as a function of reduced entangling angle on the ``bad qubit'', and see distinct improvements for smaller amounts of entangling angle in Fig. \ref{fig:badqubit}d. In particular, the less distinct clustering seen here reveals the different participatory factors of $q_{-2}$ convolved with various SU(4) merges.

\subsection{Circuit Approximation}
\label{sec:exp_circuit_approx}
Our final approach investigates the role that circuit approximation plays in improving empirical circuit performance at the expense of implementing the desired circuit exactly. Here, we use \Superstaq/ to eliminate all entangling $\zz(\theta)$ gates with an entangling angle, $\abs{\theta}$ below a threshold of 0.10 radians along with the merging of the leftover single-qubit gates remaining. Like the gate ranking investigations, all these experiments include continuously parameterized $\zz(\theta)$ and swap mirroring as default options within \Superstaq/. The selection of a threshold of 0.10 was chosen as it eliminated a substantial number of $\zz(\theta)$ gates present in the circuit $\sim15\%$, while only adding an average infidelity of $\sim0.006$ (with a maximum infidelity of $\sim0.025$) to the approximate circuit when compared to the intended circuit (see Appendix~\ref{app:stats}). Further study of the role that the threshold plays on the infidelity of the circuit (both in ideal simulation and in empirical practice) is warranted but beyond the scope of this survey of compilation optimization techniques.

\begin{figure}[h]
    \centering
    \includegraphics[width=0.45\textwidth]{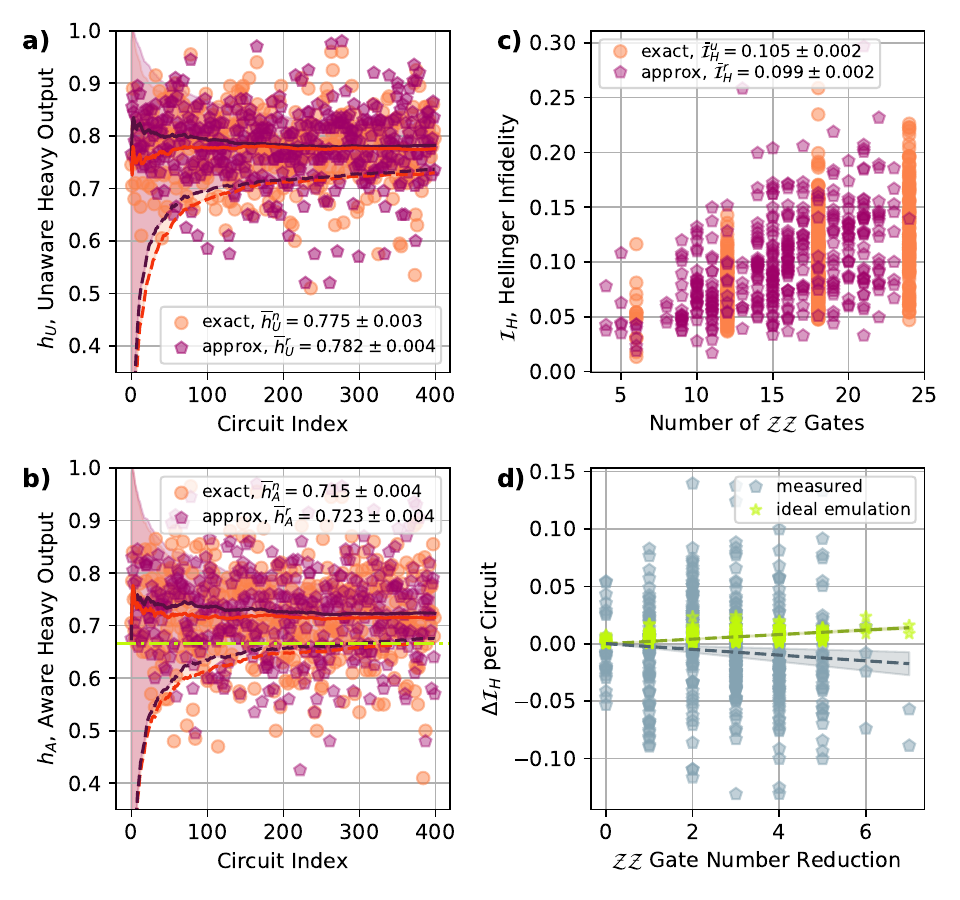}
    \caption{Comparison of 400 quantum volume circuits on a 4-qubit register compiled with continuously parameterized entangling gates, $\zz(\theta)$ with swap mirroring. Compilation using the exact circuit required (orange circles) versus approximate circuits in which entangling gates with $\abs{\theta} < 0.10$ were removed (magenta pentagons). Analysis with a) unaware heavy outputs versus run-time circuit ordering, b) aware heavy outputs versus run-time circuit ordering, c) Hellinger infidelity vs. total number of MS gates, and d) per circuit change in experimentally measured Hellinger infidelity (grey pentagons) versus the expected Hellinger infidelity increase for approximating the circuits (yellow stars) versus the overall reduction ratio of the number of MS gates. In a) and b), each running mean is presented via a solid line, while $2\sigma$ Wilson score intervals are presented with shading in between the bounds with the lower bound presented via a dashed line. In d), trend lines are presented with a 2$\sigma$ error envelope.}
    \label{fig:approx}
\end{figure}

Experimentally, we test 400 randomly selected circuits performed at 200 shots, interleaving the exact circuit and its approximation. There are distinct yet subtle improvements in the mean $\hunaware$ and $\haware$ measures presented in Fig.~\ref{fig:approx}a and Fig.~\ref{fig:approx}b. Additionally, we see both the exact and approximate circuits certify a quantum volume of $2^4$ as the $\haware$ mean and the corresponding $2\sigma$ confidence intervals are above the 2/3 threshold. Another measure to quantify the improvement from a given optimization is the number of circuits required to certify quantum volume (i.e. at what point the confidence interval crosses the threshold). Here, the approximate circuits are able certify the quantum volume measure at fewer total circuits (297 circuits) than the exact circuits require (397 circuits). Put simply, $\sim3/4$ of the number of circuits are required to certify the quantum volume benchmark when using approximate circuits. 

With the $\infidelity$ measure, we also see a small drop when approximating the circuits in Fig.~\ref{fig:approx}c. This is particularly striking as the approximated circuits are able to overcome their inherent additional infidelity incurred solely by approximation. This is more clearly seen in Fig.~\ref{fig:approx}d, where the experimentally determined per circuit $\Delta\infidelity$ is presented alongside the expected per circuit $\Delta\infidelity$ from pure simulation, both relative to the number of $\zz(\theta)$ gates by which the circuit has been reduced. (The reduction in gates is a more relevant measure than $\Gamma_{\Theta}$ which remains virtually identical per circuit on average.) There is an upward trend in the simulated results as is to be expected, while there is a noisy, yet still downward trend in the experimental results, indicating fidelity improvements gained by reducing the total number of entangling gates on average outweigh the loss in fidelity incurred by approximating the circuit.  

\section{Evaluating Stochastic and Coherent Error Contributions}
\label{sec:error_channels}
In this section, we demonstrate how the analysis approaches outlined \cref{sec:analysis_approaches} can be applied to the empirical results in \cref{sec:exp} in order to further illuminate details about the dominant errors affecting the system, and the impact of the compiler optimizations thereon.
In particular, the use of both the aware and unaware heavy outputs in tandem provides us a qualitative window into the relative roles of stochastic and coherent error sources.

We first use our experimental measurements of $\hunaware$ to estimate the relative contributions of stochastic error. For each compilation strategy, we simulate the complete set of quantum volume circuits with various simple noise models, and tune the probability of error such that the simulated $\hunaware$ matches the experimentally-measured value. This error probability is taken as an estimate of the stochastic error rate of the hardware assuming that particular error channel was dominant. By comparing these error rates across experiments we gain some insight into the impacts of each of our compiler optimizations.

We demonstrate this procedure using two different stochastic error models: gate-driven depolarization and time-based dephasing.
Note that our analysis is incapable of differentiating between these (or other) stochastic channels; the two channels are simply intended serve as representative examples indicating trends in stochastic error generally.
The depolarizing channel is implemented by inserting Pauli $X$, $Y$, or $Z$ errors with equal probability $\perr/15$ after each single qubit $\R(\theta)$ gate, and with probability $\perr/3$ per qubit after each $\zz(\theta)$ gate (reflecting an empirical observation that the error probability of single qubit gates is roughly a factor of 10 lower than that of two qubit gates).
The dephasing channel time-dependent inserts phases errors between each gate with characteristic decay rate $\perr/T_{\zz}$ (where $T_{\zz}=250$ $\mu$s is the duration of a single $\zz(\theta)$ gate).

\begin{figure}[h]
    \centering
    \includegraphics[width=0.45\textwidth]{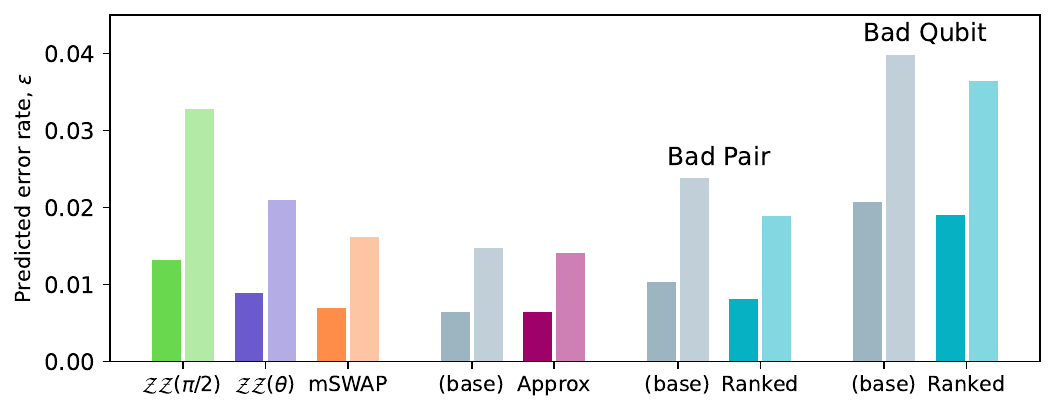}
    \caption{Simulated stochastic error rates needed to realize the empirically determined $\hunaware$ for each compilation. This assumes a single source of stochastic error, either pure depolarizing with probability of error $\perr$ (darker bar on left) or dephasing noise with characteristic decay rate $\perr/T_{\zz}$ (lighter bar on right). ``Base'' (grey) refers to the circuits without the studied optimization turned on, but with $\zz(\theta)$ and swap mirroring. Note that due to this difference in parameterization, the relative strength each error channel is not directly comparable; rather both serve to indicate trends between the different compilation methods assuming one or the other source is dominant.}
    \label{fig:err-rates-stochastic}
\end{figure}

The predicted error probabilities assuming either noise channel are shown in \cref{fig:err-rates-stochastic}.
With both models we see a clear drop in stochastic error as a result of both parameterized $\zz(\theta)$ decomposition and mirror swapping, reaffirming the $\theta$ dependence of two-qubit gate noise.
We also see a reduction in stochastic error as a result of gate-ranking in both the ``bad qubit'' and ``bad pair'' compiled circuits, demonstrating the additional stochastic error resulting from our intentional gate miscalibrations and indicating that it is successfully being circumvented by the corresponding compiler pass.
Because the approximate compilation pass removes small-angle gates but in general does not modify the remaining operations, we expect any per-gate reduction in error to be reflected in both the noisy simulations and the experimental results.
In this case we do not find an appreciable difference between the stochastic error probability predicted from our experimental results with and without circuit approximation, which is consistent with a reduction in overall circuit error due to stochastic noise.

We then repeat each simulation with the addition of various coherent errors (with the stochastic error rates fixed to that predicted from our experimental $\hunaware$). We compare these results to the experimentally-measured value of $\haware$ to predict the coherent error rate, now assuming both a single dominant coherent and single stochastic channel. Notably, we find that these coherent error estimates are largely agnostic to the chosen stochastic channel, predicting the the same coherent error rate (within $\pm1\%$) regardless of the chosen stochastic noise model. We therefore average these predictions to arrive at estimates of the coherent error rate assuming each coherent channel.

\begin{figure}[h]
    \centering
    \includegraphics[width=0.45\textwidth]{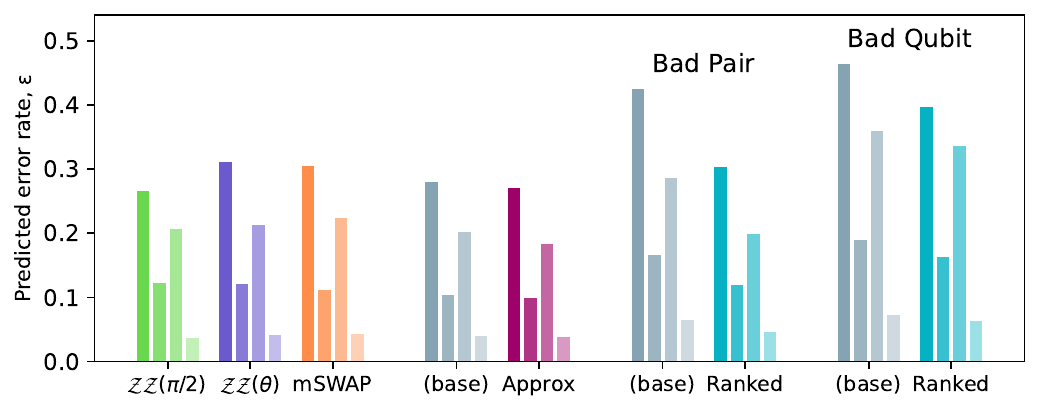}
    \caption{Estimated coherent error rate $\epsilon$ needed to realize the measured $\haware$ for each compilation, assuming the dominant source is (1, leftmost) relative over-rotation of $\zz(\theta)$ gates by a factor of $1+\epsilon$,
    (2) relative over-rotation of single-qubit $\R(\theta)$ gates by the same factor,
    (3) phase errors on single-qubit gates (excluding wrappers) proportional to their rotation angle, or by $\epsilon\theta$ radians,
    (4, rightmost) $\epsilon\pi$-radian phase errors on single-qubit wrapper gates.}
    \label{fig:err-rates-coherent}
\end{figure}

The resulting error rate predictions are shown for a variety of coherent channels in \cref{fig:err-rates-coherent}. While the true error model is likely a combination of all of these sources (and others), isolating each one can be instructive in further elucidating the dominant coherent error mechanisms in the system.
For example, we would require a 20-30\% over-rotation to account for our experimental results if that was the only source of coherent error, which is well outside the range of errors we can reasonably expect to be present on the system.
By contrast, a much smaller phase error (induced by Stark shifts that are currently uncompensated) on just the single-qubit ``wrapper'' gates implemented as part of $\zz(\theta)$ would account for the same $\haware$, making this a good candidate for additional benchmarking and calibration.

We can now examine the role the compilation optimizations have on the relative impact of coherent and stochastic error sources on the circuits.
For instance, as we switch from the $\zz(\pi/2)$ compilation to the $\zz(\theta)$ compilation, there is a significant decrease in the stochastic error rate (Fig.~\ref{fig:err-rates-stochastic}), but at the expense of a more marginal increase in the coherent error rate (Fig.~\ref{fig:err-rates-coherent}). This is expected as there should be a small increase in rotation errors in a continuously parameterized gateset (which relies on an interpolated relationship between $\theta$ and laser amplitude) over entangling gates fixed at $\theta=\pi/2$. Likewise, in the ``bad pairs'' scenario, there is a larger decrease in the coherent error rate than in the stochastic error rate, revealing the types of errors we are mitigating by this compilation are largely coherent. Thus, we have a method to see how much each compilation affects coherent and stochastic sources. It should be noted, however, that within a ``family'' of error sources (e.g. stochastic errors or coherent errors), this technique is unable to distinguish how the compilations might affect different channels with respect to one another.

In this way, our use of both the unaware and aware heavy output analyses in tandem provides us with unique insight into the relative roles of stochastic and coherent errors in our system, within the established benchmarking protocol of quantum volume.
Unfortunately, this comes with significant experimental overhead, likely making this approach impractical for larger systems.
It is therefore likely most applicable to probing smaller subsets of a larger system, in order to inform even more targeted (and costly) tomography experiments such as randomized benchmarking and its derivatives~\cite{emerson2005,dankert2009,magesan2011,magesan2012, erhard2019} (which tend to wash out coherent errors) or full gate set tomography~\cite{nielsen2021} (with extremely high experimental overhead).

\section{Conclusions}

In summary, we have developed and demonstrated a series of circuit compiler optimizations within \Superstaq/ to improve the performance of a trapped-ion processor. These compilations are tuned to the specific noise characteristics and performance of the two-qubit gates on the QSCOUT platform, leveraging the continuously parameterized two-qubit gateset as well as its inherent all-to-all connectivity. We first demonstrate a substantial improvement in circuit performance just by utilizing this continuously parameterized two-qubit gateset over a more restrictive set of maximal entanglers only. We then realize further improvements through swap mirroring which further reduces the total entangling angle per circuit.  We explore reassigning qubit roles to best utilize the highest-performing gates under two artificially constructed scenarios, one in which two orthogonal pairs perform worse than others and one in which all gate pairs associated with a single qubit perform worse. In both scenarios, we find that the circuits with knowledge of the gate fidelities outperform those without. We then demonstrate that circuit approximation to remove the least impactful two-qubit gates also can lead to marginal gains as well. Finally, we examine how different analysis approaches can elucidate the impact of these optimizations as well as estimate the role of stochastic versus coherent error channels. 

Perhaps the most conspicuous result is that these optimizations can lead to the realization of a higher quantum volume for a processor ($2^4$ in this demonstration) when using specifically tuned and optimized circuits relative to using circuits with more restrictive gatesets or those unaware of gate fidelities. Additionally, better understanding the stochastic and coherent error channels of a particular platform provides pathways to investigate and mitigate those error sources. This suite of circuit optimizations not only shows promise in the context of benchmarking like quantum volume, but other NISQ-era applications such as quantum cavity electrodynamics simulations~\cite{rubin2024}. While these particular optimizations are tuned to the performance of the QSCOUT system, they are broadly applicable to other fully connected trapped-ion processors that take advantage of continuously parameterized two-qubit gates. Similarly, this study demonstrates a general methodology that tailors circuits directly to a hardware platform and in doing so yields significant performance improvements, increasing the hardware's utility.

\section*{Acknowledgements}

We thank Joshua Goldberg and Antonio Russo for QSCOUT software support and development. This material is supported by the U.S. Department of Energy, Office of Science, Office of Advanced Scientific Computing Research under Award Number DE-SC0021526 and under its Quantum Testbed Program. Sandia National Laboratories is a multimission laboratory managed and operated by National Technology \& Engineering Solutions of Sandia, LLC, a wholly owned subsidiary of Honeywell International Inc., for the U.S. Department of Energy's National Nuclear Security Administration under contract DE-NA0003525.  This paper describes objective technical results and analysis. Any subjective views or opinions that might be expressed in the paper do not necessarily represent the views of the U.S. Department of Energy or the United States Government. SAND2024-15017O.

\appendix
\appsection{Experimental Fidelity Estimates}
\label{app:fid}
In this appendix, for completeness, we present the fidelity estimates for the $\zz$ gates for the remaining datasets (whose estimates are not detailed in the main text). These fidelity estimates are determined across each pair in the chain prior to the each measurement presented in the main text. As these are measured prior to runtime, a full fidelity estimate per pair is too time-intensive to be practical. A discussion on how actual fidelity estimates compare to these upper bound estimates on the QSCOUT system can be found in~\cite{chow2024thesis}. Table~\ref{tab:fixed_cont_mirror} presents the estimates determined prior to the experiments described in Sec.~\ref{sec:exp_cont}, which compare compilations using fixed entanglers versus those using continuously parameterized entanglers, as well as those described in Sec.~\ref{sec:exp_swap_mirror}, which compare the same compilation using continuously parameterized entanglers versus a compilation which also adds the swap mirroring optimization. 

\begin{table}[h]    
    \begin{tabular}{|c|c|c|}
        \hline
            Gate Pair & Connectivity & Est. State Fidelity \\
        \hline
        \{-1, 0\} & nearest neighbor & $0.988_{-0.005}^{+0.004}$ \\
        \hline
        \{0, 1\} & nearest neighbor & $0.984_{-0.005}^{+0.005}$ \\
        \hline
        \{1, 2\} & nearest neighbor & $0.982_{-0.006}^{+0.005}$ \\
        \hline
        \{-1, 1\} & next-nearest neighbor & $0.977_{-0.006}^{+0.005}$ \\
        \hline
        \{0, 2\} & next-nearest neighbor & $0.956_{-0.007}^{+0.007}$ \\
        \hline
        \{-1, 2\} & outer & $0.988_{-0.005}^{+0.004}$ \\
        \hline
    \end{tabular}
\caption{Fidelity estimates of two-qubit gate pairs in a four-ion chain taken prior to the measurements presented in Fig.~\ref{fig:cont} and Fig.~\ref{fig:mirrorSWAP}, comparing fixed entanglers vs. continuously parameterized entanglers vs. continuously parameterized entanglers with swap mirroring enabled. Fidelity estimates are determined in the same fashion as described in Tables~\ref{tab:bad_pairs} and~\ref{tab:bad_qubit}. Qubit name assignments can be seen in Fig.~\ref{fig:badpairs_cartoon}.}\label{tab:fixed_cont_mirror} 
\end{table}

Table~\ref{tab:approx} presents the fidelity estimates determined prior to the experiment described in Sec.~\ref{sec:exp_circuit_approx}, which compares a compilation using the exact circuit formulation versus an approximation by removal of the smallest $\zz(\theta)$ gates with $\theta<0.10$ radians.

\begin{table}[h]    
    \begin{tabular}{|c|c|c|}
        \hline
            Gate Pair & Connectivity & Est. State Fidelity \\
        \hline
        \{-1, 0\} & nearest neighbor & $0.978_{-0.006}^{+0.005}$ \\
        \hline
        \{0, 1\} & nearest neighbor & $0.984_{-0.005}^{+0.005}$ \\
        \hline
        \{1, 2\} & nearest neighbor & $0.979_{-0.006}^{+0.005}$ \\
        \hline
        \{-1, 1\} & next-nearest neighbor & $0.967_{-0.006}^{+0.006}$ \\
        \hline
        \{0, 2\} & next-nearest neighbor & $0.958_{-0.007}^{+0.007}$ \\
        \hline
        \{-1, 2\} & outer & $0.987_{-0.005}^{+0.004}$ \\
        \hline
    \end{tabular}
\caption{Fidelity estimates of two-qubit gate pairs in a four-ion chain taken prior to the measurements presented in Fig.~\ref{fig:approx}, comparing exact circuit formulations vs. approximate circuit formulations. Fidelity estimates are determined in the same fashion as described in Tables~\ref{tab:bad_pairs} and~\ref{tab:bad_qubit}. Qubit name assignments can be seen in Fig.~\ref{fig:badpairs_cartoon}.}\label{tab:approx} 
\end{table}

\appsection{Compilation Optimization Statistics}
\label{app:stats}

In this appendix, we present further analysis of the circuits being performed for some of the compilation optimizations studied. Additionally, we present further statistical analyses on all of the data collected from the experimental apparatus (corresponding to all compilations studied). We begin by analyzing the circuits themselves to better understand when and how a particular compilation affects each particular randomly generated circuit.

\begin{figure}[h]
    \centering
    \includegraphics[width=0.45\textwidth]{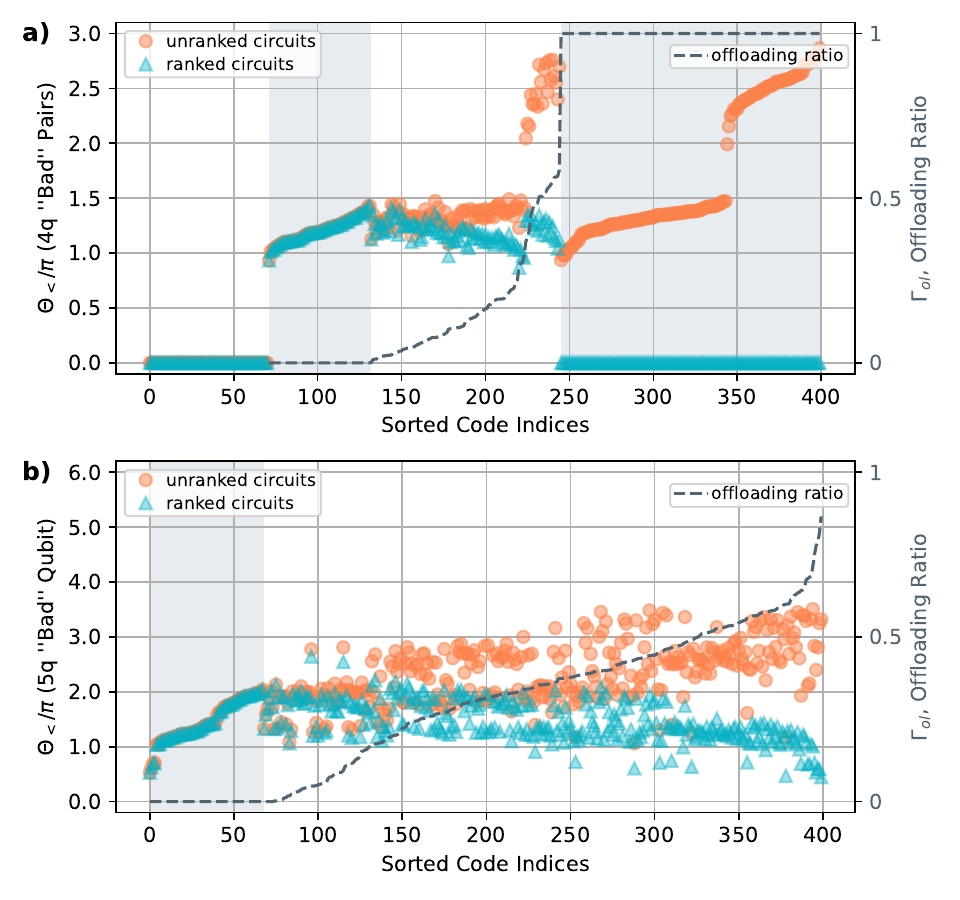}
    \caption{Per circuit analysis of total entangling angle, $\Theta$ on the a) orthogonal ``bad pairs'' in a 4-qubit register and b) the ``bad qubit'' in a 5-qubit register. The original circuits (orange circles) and ranked circuits (teal triangles) are sorted to reveal different groups of the effectiveness of the ranking compilation and shaded accordingly. In a), the first 71 circuits have no participation of the ``bad pairs'' regardless. The next 61 are circuits in which participation of the ``bad pairs'' already has the lowest possible $\Theta$. The following 113 reveal circuits where partial offloading is achieved, the remaining 155 indicate circuits in which complete offloading of $\Theta$ is realized. In b), the first 67 circuits are those in which the participation of the ``bad qubit'' is already minimized. The remaining 333 are circuits where partial offloading occurs. The dashed grey line indicates the offloading ratio per circuit and corresponds to the right axis in both plots.}
    \label{fig:ranking_stats}
\end{figure}

For the ``bad pairs'' scenario in which two orthogonal pairs in a four-qubit register have diminished performance (as described in Sec.~\ref{sec:bad_pairs}), we find four distinct possibilities for how the compilation affects any given circuit (Fig.~\ref{fig:ranking_stats}a) across all 400 circuits used. First, there are 71 circuits (17.75\%) that have no participation from those pairs to begin with and are unaffected. Another 61 circuits (15.25\%) are unaffected by this compilation as all pairs are involved in the circuit and those two orthogonal pairs are already participating with the the least entangling angle possible. 113 circuits (28.25\%) partially offload participation of the orthogonal pairs, while the remaining 155 circuits (38.75\%) offload participation of those two pairs completely. Thus, at this register size of four qubits, we find this compilation was effective for $\sim2/3$ of the circuits. We define an offloading ratio, $\Gamma_{ol}$ (similar to the reduction ratio $\Gamma_{\Theta}$ defined in Eq.~\ref{eq:reduction_ratio}), which is the reduction in $\Theta_{<}$ between the unranked and ranked circuits as compared to the total $\Theta_{<}$ in the unranked circuit, or:
\begin{equation}
\Gamma_{ol} =\frac{\Theta_{<,i} - \Theta_{<,j}}{\Theta_{<,i}}
\end{equation}
in which $i$, $j$ refer to the unranked and ranked circuits respectively. This is plotted in Fig.~\ref{fig:ranking_stats} with a dashed line referenced to the right axis.

With the ``bad qubit'' scenario in the five-qubit register (Sec.~\ref{sec:bad_qubit}), the optimizer works to relabel the qubit to reduce its entangling angle largely based on how many layers it participates in. In the 400 circuits tested, this leads to a partial offloading on 333 circuits ($83.25\%$) and the remainder are identical to the original circuit. In Fig.~\ref{fig:ranking_stats}b, the overall entangling angle on the ``bad qubit'' is presented for both the ranked and unranked 400 circuits. In particular, there are no cases of complete offloading, so all circuits are still affected by the poorer performing gates to some degree.

\begin{figure}[h!!!!!!!!!!!]
    \centering
    \includegraphics[width=0.45\textwidth]{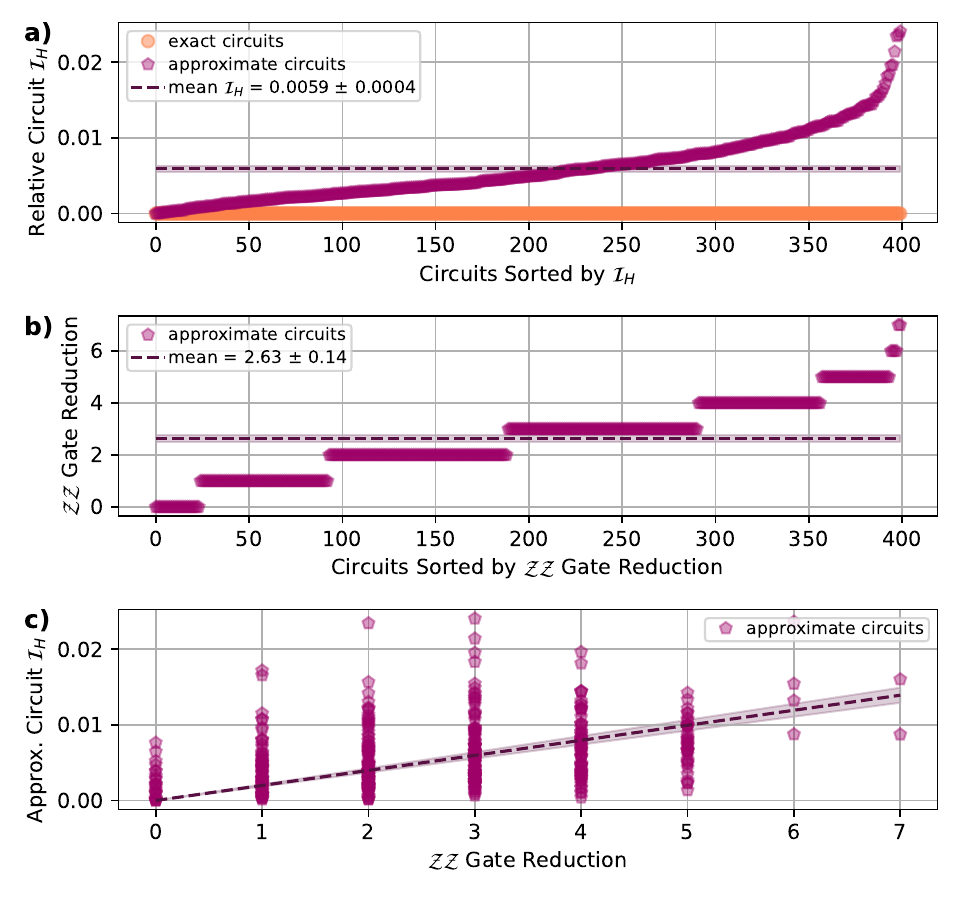}
    \caption{Per circuit analysis of the approximate circuits. a) Error-free simulation of the natural infidelity of the exact circuits (orange circles) and the approximate circuits (magenta pentagons) are plotted in which the circuits are sorted by increasing infidelity, along with a mean line (and 2$\sigma$ standard error on the mean) indicating the average infidelity per approximate circuit. b) The number of $\zz$ gates by which each approximate circuit is reduced are plotted with the circuits sorted by increasing $\zz$ gate reduction. A mean line and 2$\sigma$ standard error on the mean are also plotted. c) The natural infidelity of the approximate circuits is plotted versus the $\zz$ gate reduction. A trend line is presented with 2$\sigma$ error.}
    \label{fig:approx_stats}
\end{figure}

In the case of the circuit approximation, we examine the natural circuit infidelities that arise from this particular approximation: removing all $\zz(\theta)$ gates with $\abs{\theta}<0.10$ for a four-qubit quantum volume circuit. Simulating the circuits, we find the average $\infidelity$ is $\sim0.006$ with a maximal $\infidelity$ of $\sim0.025$ as shown in Fig~\ref{fig:approx_stats}a. Examining the circuits more closely, we find that on average $\sim2.6~\zz$ gates are removed per circuit, with a maximal removal of 7 gates in total for 2 of the 400 circuits, in Fig.~\ref{fig:approx_stats}b. As is to be expected, there is also rough trend for \emph{decreasing} performance (i.e. increasing $\infidelity$) as a function of increasing $\zz$ gate reduction in Fig.~\ref{fig:approx_stats}c. However, it is important to note that the empirical results in Fig.~\ref{fig:approx}d show instead a rough trend for a \emph{gain} in performance (i.e. decreasing infidelity), suggesting that a reduction in $\zz$ gates via this approximation in general is able to outperform the loss of fidelity that is incurred by virtue of the approximation.

\begin{table*}[t]    
    \begin{tabular}{|c|c|c|c|c|c|c|}
        \hline
         Comparison & \multicolumn{2}{c|}{$\hunaware$} & \multicolumn{2}{c|}{$\haware$} & \multicolumn{2}{c|}{$\infidelity$} \\
        \hline
        & t-statistic & p-value & t-statistic & p-value & t-statistic & p-value \\
        \hline
        $\zz(\pi/2)$ vs. $\zz(\theta)$ & \cellcolor{teal!20} -4.679 & \cellcolor{teal!20} 0.000 & \cellcolor{teal!20} \cellcolor{teal!20} -5.810 & \cellcolor{teal!20} 0.000 & \cellcolor{teal!20} 6.740 & \cellcolor{teal!20} 0.000 \\
        \hline
        $\zz(\theta)$ vs. +mSWAP & \cellcolor{teal!20} -2.284 & \cellcolor{teal!20} 0.023 & -1.609 & 0.108 & 1.407 & 0.160 \\
        \hline
        unranked vs. ranked ``bad pairs'' & \cellcolor{teal!20} -2.934 & \cellcolor{teal!20} 0.003 & \cellcolor{teal!20} -6.647 & \cellcolor{teal!20} 0.000 & \cellcolor{teal!20} 8.419 & \cellcolor{teal!20} 0.000 \\
        \hline
        unranked vs. ranked ``bad qubit'' & -1.574 & 0.116 & \cellcolor{teal!20} -2.232 & \cellcolor{teal!20} 0.026 & 1.145 & 0.253 \\
        \hline
        exact vs. approx & -1.330 & 0.184 & -1.354 & 0.176 & \cellcolor{teal!20} 2.003 & \cellcolor{teal!20} 0.046 \\
        \hline
    \end{tabular}
\caption{Further statistics to compare compilation optimizations. To determine the likelihood the measured distributions are drawn from the same distribution, the t-statistic and p-value for each compilation optimization and analysis method are presented. Highlighted in teal are values which pass the typical threshold to determine the statistical significance that two distributions are distinct from one another. Thresholds are $>\lvert2\rvert$ for the t-statistic and $<0.05$ for the p-value.}\label{tab:stats} 
\end{table*}

Next, in Table~\ref{tab:stats}, we process the collected empirical data to compute a t-statistic and p-value for all comparisons presented in the main text across all analysis approaches: $\hunaware$, $\haware$, and $\infidelity$. While there are certainly more complex (and more accurate) ways to determine the likelihood the analyzed empirical distributions are drawn from the same distribution, this is a rudimentary and common method. It should be noted that while some t-statistics (and p-values) \emph{do not} fall above (below) typical thresholds of $\lvert2\rvert$ (0.05), for each comparison we make, there is at least one analysis approach which yields a t-statistic (and p-value) which \emph{does} lie within the generally accepted bounds for determining distinctness, suggesting all compilation optimizations provide a distinct change to the performance of the system.

\begin{table}[h]    
    \begin{tabular}{|c|c|c|c|c|}
        \hline
         Compilation & \# of Circuits & Compilation & \# of Circuits & Figure \\
        \hline
        $\zz(\pi/2)$ & -- & $\zz(\theta)$ & \emph{1536} & Fig.~\ref{fig:cont} \\
        \hline
        $\zz(\theta)$ & \emph{1536} & +mSWAP & \emph{592} & Fig.~\ref{fig:mirrorSWAP} \\
        \hline
        unranked & -- & ``bad pairs'' & \emph{1057} & Fig.~\ref{fig:badpairs} \\
        \hline
        unranked & -- & ``bad qubit'' & -- & Fig.~\ref{fig:badqubit} \\
        \hline
        exact & 397 & approximate & 297 & Fig.~\ref{fig:approx} \\
        \hline
    \end{tabular}
\caption{Quantum Volume Certification. Number of circuits required to certify quantum volume of $2^4$ (or $2^5$ in the case of Fig.~\ref{fig:badqubit}) for all compilations presented in the main text. For distributions with means below 2/3, no quantum volume certification can be achieved. For distributions with means above 2/3 but CI below 2/3, the extrapolated number of circuits required to certify is presented in italics. For distributions with means and CI above 2/3, the measured number of circuits is presented.}\label{tab:wilson_score_crosses} 
\end{table}

Finally, in Table~\ref{tab:wilson_score_crosses}, we examine the certification of quantum volume based on the distributions measured in the main text. While there are some slight differences in how quantum volume certification is defined across the literature~\cite{cross2019, baldwin2022qv, pelofske2022}, it in general occurs when the lower bound $2\sigma$ confidence interval (CI) crosses the 2/3 threshold. In our case, we use one of the more conservative approaches to defining CI, with the Wilson score interval~\cite{wilson1927} with continuity correction defined as,
\begin{multline}
    w^{-} = \mathrm{max}\Biggl\{0,\ldots\\
    \left.\frac{2N\hat{p} + z^2 - z\sqrt{z^2 - 1/N + 4N\hat{p}(1-\hat{p}) + (4\hat{p} -2)} - 1}{2(N+z^2)}\right\},
\end{multline}\label{eq:wilson}
where $N$ is the number of random circuits evaluated, $z=2$ defines the $2\sigma$ error level, and $\hat{p}$ is the mean of the distribution. 

We then evaluate the experimental results for the different compilation optimizations presented in the main text. For results in which the mean lies below the 2/3 threshold, no certification is possible. In several cases, the mean lies above the 2/3 threhsold, but for the limited number of circuits tested (200 or 400), its empirically determined CI still lies below the threshold. As such, using the experimentally determined mean, we extrapolate the number of circuits required to have a CI that lies above the threshold based on Eq.~\ref{eq:wilson}. In particular, when comparing the $\zz(\theta)$ compilations with and without swap mirroring (Fig.~\ref{fig:cont}), we find that the mean of the circuits with swap mirroring would only require 592 circuits, while the same circuits without would require 1536 circuits to be evaluated. Our final demonstration looking at exact and approximate circuits showed the CI did cross the 2/3 threshold for both measurements given the number of circuits evaluated. Here, the approximate circuits reach certification at 297 circuits, while the exact circuits require 397 circuits. 

We note that a $2\sigma$ Wald interval~\cite{cross2019} would certify these optimizations at fewer circuits. For example, the approximate circuits are certified at 234 circuits with the Wald interval rather than 297 circuits with the its corresponding Wilson score interval. Likewise, a bootstrapping method which also takes into account \emph{the number of shots} rather than just the number of distinct circuits in each measurement, such as the one presented in~\cite{baldwin2022qv}, would likely certify at even fewer circuits.

\newpage
\bibliography{refs}

\end{document}